\documentclass[aps, pre, twocolumn, superscriptaddress, 
preprintnumbers, amsmath, amssymb]{revtex4}
\usepackage{graphicx}	
\graphicspath{ {./images/} }
\usepackage{bm}		
\usepackage{color}	
\usepackage{dcolumn}    
\usepackage{multirow}   
\usepackage{longtable,epsfig}

\usepackage{txfonts}
\DeclareMathAlphabet{\mathcal}{OMS}{cmsy}{m}{n}
\DeclareSymbolFont{largesymbols}{OMX}{cmex}{m}{n}

\usepackage[
            pdfstartview=FitH,
            CJKbookmarks=true,
            bookmarksnumbered=true,
            bookmarksopen=true,
            colorlinks,
            linkcolor=blue,
            anchorcolor=blue,
            citecolor=blue
            ]{hyperref}

\begin{document}

\title{Negative-parity high-spin structure of $^{105}$Pd}

\author{B.~Kruzsicz}
\affiliation{HUN-REN Institute for Nuclear Research, Hungarian Academy of Sciences, Pf. 51, 4001 Debrecen, Hungary}
\affiliation{University of Debrecen, Doctoral School of Physics, 4032 Debrecen, Egyetem t\'er 1,
Hungary}

\author{D.~Sohler}
\email [Corresponding author: ] {sohler@atomki.hu}
\affiliation{HUN-REN Institute for Nuclear Research, Hungarian Academy of Sciences, Pf. 51, 4001 Debrecen, Hungary}

\author {J.~Tim\'ar}
\affiliation{HUN-REN Institute for Nuclear Research, Hungarian Academy of Sciences, Pf. 51, 4001 Debrecen, Hungary}

\author{I.~Kuti}
\affiliation{HUN-REN Institute for Nuclear Research, Hungarian Academy of Sciences, Pf. 51, 4001 Debrecen, Hungary}

\author{Q.~B.~Chen}
\affiliation{Department of Physics, East China Normal University, Shanghai 200241, China}

\author{S.~Q.~Zhang}
\affiliation{State Key Laboratory of Physics and Technology, School of Physics, Peking University, Beijing 100871, China}

\author{J.~Meng}
\affiliation{State Key Laboratory of Physics and Technology, School of Physics, Peking University, Beijing 100871, China}

\author{P.~Joshi}
\affiliation{School of Physics, Engineering and Technology, University of York, York, YO10 5DD, United Kingdom}

\author{R.~Wadsworth}
\affiliation{School of Physics, Engineering and Technology, University of York, York, YO10 5DD, United Kingdom}

\author{K.~Starosta}
\affiliation{Department of Chemistry, Simon Fraser University, Burnaby, British Columbia V5A 1S6, Canada}

\author{A.~Algora}
\affiliation{HUN-REN Institute for Nuclear Research, Hungarian Academy of Sciences, Pf. 51, 4001 Debrecen, Hungary}
\affiliation{Instituto de Fisica Corpuscular, CSIC-University of Valencia, E-46071, Valencia, Spain}

\author{P.~Bednarczyk}
\affiliation{Institute of Nuclear Physics Polish Academy of Sciences, PL-31342 Krakow, Poland}

\author{D.~Curien}
\affiliation{Universit\'e de Strasbourg, CNRS, IPHC UMR 7178, 67037 Strasbourg, France}

\author{Zs.~Dombr\'adi}
\affiliation{HUN-REN Institute for Nuclear Research, Hungarian Academy of Sciences, Pf. 51, 4001 Debrecen, Hungary}

\author{G.~Duch\^ene}
\affiliation{Universit\'e de Strasbourg, CNRS, IPHC UMR 7178, 67037 Strasbourg, France}

\author{A.~Gizon}
\affiliation{LPSC, IN2P3-CNRS/UJF, F-38026 Grenoble-Cedex, France}

\author{J.~Gizon}
\affiliation{LPSC, IN2P3-CNRS/UJF, F-38026 Grenoble-Cedex, France}

\author{D.~G.~Jenkins}
\affiliation{School of Physics, Engineering and Technology, University of York, York, YO10 5DD, United Kingdom}

\author{T.~Koike}
\affiliation{Graduate School of Science, Tohoku University, Sendai, 980-8578, Japan}

\author{A.~Krakó}
\affiliation{HUN-REN Institute for Nuclear Research, Hungarian Academy of Sciences, Pf. 51, 4001 Debrecen, Hungary}
\affiliation{University of Debrecen, Doctoral School of Physics, 4032 Debrecen, Egyetem t\'er 1,
Hungary}

\author{A.~Krasznahorkay}
\affiliation{HUN-REN Institute for Nuclear Research, Hungarian Academy of Sciences, Pf. 51, 4001 Debrecen, Hungary}

\author{J.~Moln\'ar}
\affiliation{HUN-REN Institute for Nuclear Research, Hungarian Academy of Sciences, Pf. 51, 4001 Debrecen, Hungary}

\author{B.~M.~Nyak\'o}
\affiliation{HUN-REN Institute for Nuclear Research, Hungarian Academy of Sciences, Pf. 51, 4001 Debrecen, Hungary}

\author{E.~S.~Paul}
\affiliation{Oliver Lodge Laboratory, Department of Physics, University of Liverpool, Liverpool L69 7ZE, United Kingdom}

\author{G.~Rainovski}
\affiliation{Faculty of Physics, St. Kliment Ohridski University of Sofia, 1164 Sofia, Bulgaria}

\author{J.~N.~Scheurer}
\affiliation{Universit\'e Bordeaux 1, IN2P3- CENBG - Le Haut-Vigneau BP120 33175, Gradignan Cedex, France}

\author{A.~J.~Simons}
\affiliation{School of Physics, Engineering and Technology, University of York, York, YO10 5DD, United Kingdom}

\author{C.~Vaman}
\affiliation{Department of Physics and Astronomy, SUNY, Stony Brook,~New York,~11794-3800, USA}

\author{L.~Zolnai}
\affiliation{HUN-REN Institute for Nuclear Research, Hungarian Academy of Sciences, Pf. 51, 4001 Debrecen, Hungary}

\date{\today}   

\begin{abstract}

Negative-parity medium- and high-spin structure of the nucleus 
$^{105}$Pd was studied through the $^{96}\textrm{Zr}(^{13}\textrm{C}, 4n)^{105}\textrm{Pd}$ 
reaction at incident energies of 51 and 58~MeV, using the EUROBALL IV 
$\gamma$-ray spectrometer in conjunction with the DIAMANT charged particle 
array. New bands have been observed and the previously reported bands 
have been extended to higher energies and spins. Altogether six 
decoupled bands with $E2$ transitions and 
one strongly coupled band with $M1+E2$ transitions 
have been observed. The observed energy spectra and $B(M1)/B(E2)$ 
ratios are compared with results of quantum particle rotor model 
calculations. Based on these comparisons, quasiparticle configurations 
can be assigned to two newly observed decoupled bands as well 
as to the strongly coupled band. The previously emerged possible 
interpretation for the third decoupled band as a two-phonon wobbling 
excitation lacks support. The observations indicate possible $\gamma$-band 
nature for this band. The strongly coupled band, consistently with the 
absence of another observed strongly coupled band in this experiment, 
does not exhibit chirality.

\end{abstract}


\maketitle

\section{Introduction}
Transitional nuclei in the $A\sim100$ mass region are known to have triaxial shapes
leading to the appearance of exotic phenomena such as the chiral twin bands and the nuclear
wobbling motion. Indeed, the appearance of chiral twin bands has been observed in several odd-odd 
and odd-mass Rh and Ag nuclei around $^{105}$Pd, and nuclear wobbling motion has been reported
recently in $^{105}$Pd, confirming the triaxially deformed shape of these nuclei.

Nuclear chirality was first predicted by Frauendorf and Meng~\cite{1}. It was shown that in
chiral geometry the total angular momentum vector of the rotating triaxial nucleus lies outside
the three principal planes, generating a pair of $\Delta I=1$ nearly degenerate bands with the
same parity. To establish the chiral geometry, particle-type and hole-type high-$j$ quasiparticles
are needed to be coupled to the triaxial core. In the $A\sim100$ mass region the $\nu(h_{11/2})$ 
and the $\pi(g_{9/2})$ states provide the particle-type and the hole-type high-$j$ quasiparticles, 
respectively. 

In the $A\sim 100$ mass region, chirality has been observed first 
in the odd-odd $^{104}$Rh~\cite{104rh}, with 
the $\pi(g_{9/2})\otimes \nu(h_{11/2})$ configuration. 
Later, it has also been reported in other 
nearby odd-odd Rh~\cite{102rh,106rh} and Ag~\cite{104ag,106ag} nuclei. 
Composite chiral configurations, containing more than one unpaired neutron, have also been 
observed in the odd-proton odd-mass nuclei in this region with configurations of 
$\pi(g_{9/2})\otimes \nu(h_{11/2})^2$~\cite{105rh,103rh} and of
${\pi}g_{9/2}{\otimes}{\nu}h_{11/2}(g_{7/2}, d_{5/2})$~\cite{103rh2}. Evolution of the chiral rotation 
mode in these rhodium isotopes has been described in Ref.~\cite{evol}. However, no chiral bands 
have been observed up to now in the odd-neutron odd-mass nuclei in this region yet, although the 
$\nu(h_{11/2})$ and the $\pi(g_{9/2})$ states are active valence states in these nuclei, too. 
Therefore, it is interesting to search for and study bands with configurations involving these states
experimentally in the Pd nuclei. The expected such configurations in these nuclei are the 
$\pi(g_{9/2})^2\otimes \nu(h_{11/2})$ negative-parity, and the $\pi(g_{9/2}, p_{1/2})\otimes \nu(h_{11/2})$ positive-parity configurations.

The name of wobbling was introduced by Bohr and Mottelson~\cite{Bohr1975book}
for an approximation of the complex three-dimensional rotation 
of triaxially deformed nuclei. This approximation is described 
by a one-dimensional rotation around the principal axis having 
the largest moment of inertia with harmonic oscillation of the 
rotation axis around the angular momentum vector. This type of 
motion is characterized by a series of rotational $E2$ bands 
corresponding to the different oscillation quanta ($n$), and by 
large $B(E2)$ values between the successive bands.
In the past decade, several candidate rotational bands have been
observed for this phenomenon. The first such candidate 
was reported in the triaxial strongly deformed bands 
of the $^{163}$Lu nucleus~\cite{163Lu,163Lu2}, 
where both the one-phonon ($n = 1$) and the two-phonon ($n = 2$) 
wobbling bands have been observed. Later many other candidates 
in medium- or high-spin bands of different mass regions have 
been reported. The theoretical interpretations of 
these bands have been successfully carried out using the 
quasiparticle-plus-triaxial-rotor model. With some additional 
assumptions this model was even able to explain the observed
differences of the experimental bands from the Bohr and Mottelson
type predictions~\cite{transverse}. Recently, another 
interpretation called as tilted precession (TiP) was 
proposed~\cite{Lawrie}. This interpretation argues that 
the solutions from the quasiparticle-plus-triaxial-rotor model 
lack the quantization (typically observed in ideal phonon 
excitations) for both excitation energies and transition 
probabilities. Nevertheless, the interpretation on the 
wobbling candidates remain under debate. More detailed 
discussions on this can be found in the monograph~\cite{Petrache2024book}.

In the $A \sim 100$ mass region, wobbling motion has only been reported 
in $^{105}$Pd~\cite{105Pd}. In this nucleus only the one-phonon 
wobbling band has been observed to date, with the 
negative-parity $h_{11/2}$ configuration. As the second wobbling 
band is not known, the quantization criteria could not be examined
experimentally. Theoretically, more odd-neutron wobbling candidates 
were predicted around this nucleus~\cite{H.M.Dai2023PRC}.
In the present work, we have studied the negative-parity
medium- and high-spin bands of $^{105}$Pd based on a high-statistics 
EUROBALL IV experiment in order to search for the possible two-phonon 
wobbling band and also for chiral bands. Part of the results has 
already been published in Refs.~\cite{105Pd, 105Pd-cp}. In this work, we have extended the analysis to all the observed negative-parity bands.

\section{Experimental methods and results}

Excited states in $^{105}$Pd were populated using the 
$^{96}\textrm{Zr}(^{13}\textrm{C}, 4n)^{105}\textrm{Pd}$ 
fusion evaporation reaction at IPHC, Strasbourg. The beam 
was provided by the Vivitron accelerator at energies of 51 
and 58~MeV. The $^{13}$C beam impinged on a stack of two 
metallic foil targets, each having thickness of 
$\sim$$0.6~\textrm{mg}/\textrm{cm}^{2}$, 
and being enriched to $86\%$ in $^{96}$Zr. The emitted 
$\gamma$ rays were detected by the EUROBALL IV 
spectrometer~\cite{EB4} equipped with 15 Cluster detectors 
at backward angles and 24 Clover detectors at 90$^{\circ}$ 
relative to the beam direction. In order to eliminate the 
contaminant reaction channels produced by evaporation of 
protons or $\alpha$ particles from the $^{109}$Pd compound 
nuclei, $\gamma$ rays were measured in coincidence with 
light charged particles. The charged particles were 
detected by the highly efficient DIAMANT array~\cite{jns, gal}, 
which consisted of 88 CsI detector elements. It has been 
used as an off-line veto to suppress the contaminants. 
A total of $\sim 2\times 10^9$ triple- and higher-fold 
coincidence events were obtained and stored onto magnetic 
tapes. Of which, $\sim 7 \times 10^8$ events belonged 
to the $^{105}$Pd channel. A $^{152}$Eu source, placed 
at the target position, was used for energy and efficiency 
calibration of the Ge detectors. Systematic errors for 
these calibrations have been estimated to be 0.2-0.3~keV 
and $\sim 5\%$, respectively. 

\begin{figure}
\centering
\includegraphics[width=7.0 cm,angle=0,bb=70 30 670 620]{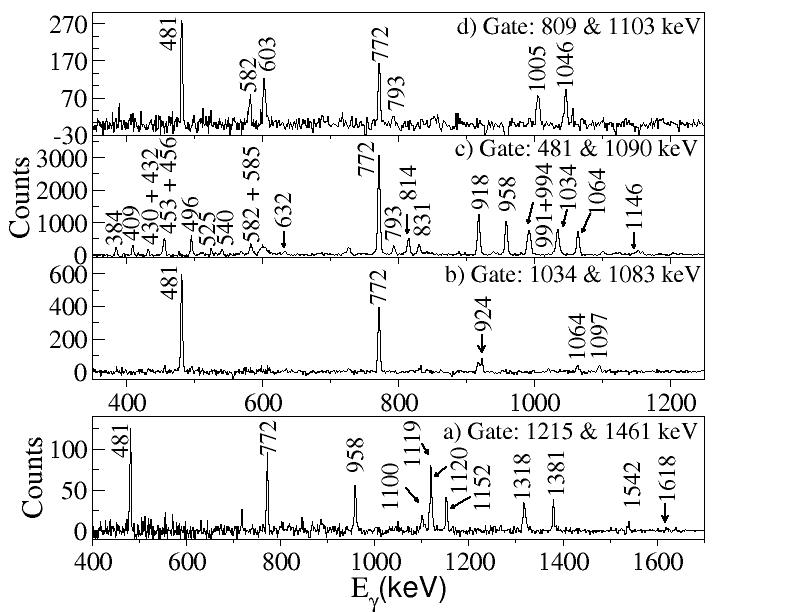}
\caption{Typical double $\gamma$-ray gated, background 
subtracted, $\gamma\gamma\gamma$-coincidence spectra 
a, b, c, and d, showing the placement of bands N2, N1, N3, and N4,
respectively. The double $\gamma$-ray gates used are indicated on the panels.}
\label{gate}
\end{figure}

\begin{figure}
\centering
\includegraphics[width=7.0 cm,angle=0,bb=80 0 650 550]{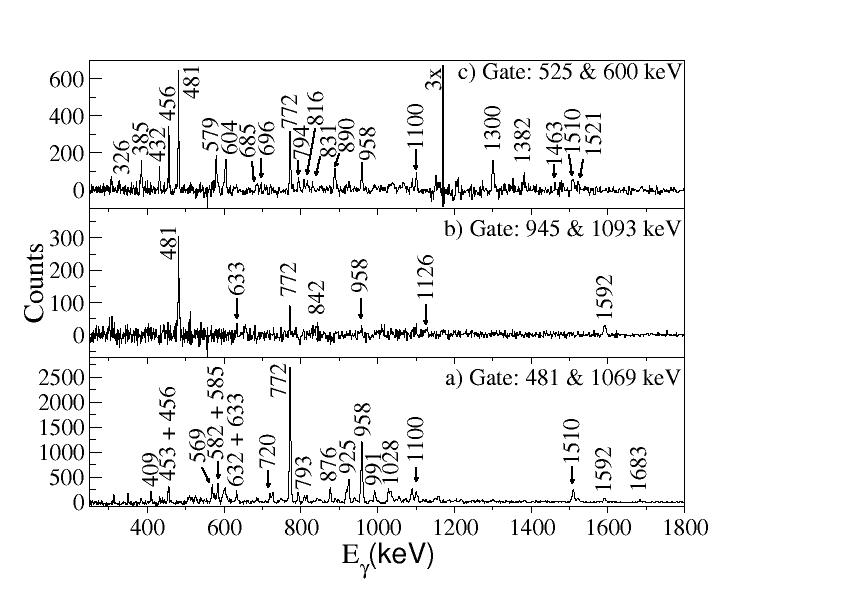}
\vspace{-0.5cm}
\caption{Typical double $\gamma$-ray gated, background subtracted, 
$\gamma\gamma\gamma$-coincidence spectra a, b, and c, showing 
the placement of bands N5, N6, and N7, respectively. 
The double $\gamma$-ray gates used are indicated on the panels.}
\label{gate2}
\end{figure}

The data received from the Ge detectors were sorted offline into two- and three-dimensional histograms with the condition of nondetection of any charged particle. The level scheme of $^{105}$Pd has been constructed with the RADWARE software package~\cite{rad} using the triple coincidence relations in the three-dimensional histograms. Energy and intensity balances of the observed $\gamma$ rays have also been taken into account. Several new $\gamma$ rays and rotational bands have been observed in the studied nucleus. The energies and relative intensities of the observed $\gamma$ rays are given in Table~\ref{table1}. The $\gamma\gamma\gamma$-coincidence spectra (Fig~\ref{gate} and Fig~\ref{gate2}) show the evidence for the placements of the observed $\gamma$-rays. Figure~\ref{gate} shows the bands N2, N1, N3 and N4, while Figure~\ref{gate2} shows the bands N5, N6, N7. In both figures double gates were used on the panels. Altogether 150 transitions have been placed into the negative-parity level scheme of $^{105}$Pd in the present work, two third of which have been assigned to the studied nucleus for the first time.

Information on the multipolarity of transitions with large enough intensity was extracted from an analysis of directional correlation of oriented nuclei (DCO) ratios~\cite{dco}. For the DCO analysis, data obtained from the cluster detectors mounted at an average angle of $156^{\rm o}$ and the clover detectors arranged at about $90^{\rm o}$ were used. A non-symmetric $E_{\gamma}$-$E_{\gamma}$ 
matrix was created, comprising the coincident $\gamma$ rays observed in the cluster detectors along one axis and the $\gamma$ rays observed in the clover detectors along the other axis. The coincidence-intensity ratio $R_{\rm DCO}$=$I_{\gamma\gamma}(156^{\circ},90^{\circ}$[gate])/$I_{\gamma\gamma}(90^{\circ},156^{\circ}$[gate]) was extracted from this matrix applying standard gating procedures and corrections for the different efficiencies of the clover and
the cluster detector rings. Expected DCO ratios have also been calculated for the experimental geometry as described in Ref.~\cite{105rh}. The attenuation coefficients of incomplete alignment were fitted to strong transitions with known multipolarities in the calculations. These estimates revealed that a value of $R_{\rm DCO}=1.0$ is expected for a pure stretched quadrupole transition 
and $\approx 0.6$ for a pure stretched dipole one, when the gating transition is a stretched quadrupole $\gamma$ ray. With this gating condition $R_{\rm DCO}$ for a pure non-stretched dipole transition is approximately the same as for a stretched quadrupole transition. For mixed $M1+E2$ transitions $R_{\rm DCO}$ ratios can vary between 0.3 and 1.2 depending on the $\delta$ mixing ratio of the $\gamma$ 
ray. Altogether we have deduced DCO ratios for $\sim 55\%$ of the transitions. The obtained $R_{\rm DCO}$ ratios are summarized in Table~\ref{table1}. In all cases, the gating transition was a stretched quadrupole $\gamma$ ray.

The multipolarity assignments were further corroborated by extracting the electromagnetic character of the transitions by measuring the linear polarization of the $\gamma$ rays. For this purpose, the four-element clover detectors placed close to $90^{\rm o}$ relative to the beam direction were used as Compton polarimeters~\cite{polD, pol1, ru15}. Two matrices were constructed from $\gamma\gamma$-events; single hits in any detectors were placed on one axis while the added-back double-hit scattering events were placed on the other axis. In the first matrix the scattering events took place perpendicular, while in the second matrix parallel to the reaction plane. The number of perpendicular ($N_{\perp}$) and parallel ($N_{\parallel}$) scatters for a given $\gamma$ ray were obtained from spectra gated on the single-hit axis of the respective matrix by transitions in coincidence with the given $\gamma$ ray. Assuming that each clover crystal has equal efficiency, an experimental linear polarization is defined as
\begin{equation}
P=\frac{1}{Q}\frac{N_{\perp}-N_{\parallel}}{N_{\perp}+N_{\parallel}},
\end{equation}
where $Q$ is the polarization sensitivity for the clover detectors, which is a function of the $\gamma$-ray energy~\cite{polD, pol1, ru15}. $N_{\perp}$ and $N_{\parallel}$ denote the number of events scattered perpendicular and parallel to the reaction plane, respectively. One notes that $P>0$ is characteristic for stretched $E1$, $E2$, and non-stretched $M1$ transitions, while $P<0$ characterizes the stretched $M1$ and non-stretched $E1$ transitions. Linear polarization has been obtained for $\sim 40\%$ of the transitions. The results of the linear polarization analysis are summarized in Table~\ref{table1}.

During the multipolarity determination we have considered stretched $M1$, stretched $E2$, and non-stretched $\Delta I=0$ $M1$ transitions based on the DCO and linear polarization results. Firm $M1$ multipolarity has been assigned to transitions, when supported by linear polarization results. Otherwise dipole assignments have been proposed. If only a DCO ratio could be deduced for a $\gamma$ ray with quadrupole character, $E2$ multipolarity has been suggested for it. Our spin-parity assignments are based on the multipolarities obtained, on the decay pattern of the excited states and on the assumption that in a fusion-evaporation reaction the spins are increasing with increasing excitation energies. The deduced multipolarities of $\gamma$ rays together with the spin-parity values of the initial states are shown in Table~\ref{table1}.

\LTcapwidth=\textwidth
\setlength{\tabcolsep}{7.5pt}
\renewcommand{\arraystretch}{1.3}
\begin{longtable*}{@{\extracolsep{\fill}}llllllrllrll @{\extracolsep{\fill}}}
\caption{Energies, relative intensities, DCO ratios, linear polarizations and deduced multipolarities of transitions assigned to $^{105}$Pd in the present work, as well as energies of their initial states, band labels and spin-parities of their initial and final states.} 
\label{table1}
\\
\hline
\hline
\multicolumn{1}{l}{$E_{\gamma}$ (keV)} &
\multicolumn{1}{l}{$I_{\gamma}$ (rel.)} &
\multicolumn{1}{l}{$R_{\rm DCO}$} &
\multicolumn{1}{l}{$P$} &
\multicolumn{1}{l}{Mult.} &
\multicolumn{1}{l}{$E_i$ (keV)} &
\multicolumn{1}{r}{Band$_i$}&
\multicolumn{1}{l}{$\rightarrow	$}&
\multicolumn{1}{l}{Band$_f$}&
\multicolumn{1}{r}{$I_i^\pi$}&
\multicolumn{1}{l}{$\rightarrow	$}&
\multicolumn{1}{l}{$I_f^\pi$}\\
\hline
\endfirsthead
\caption{(\textit{Continued.})}
\\
\hline
\hline
\multicolumn{1}{l}{$E_\gamma$ (keV)} &
\multicolumn{1}{l}{$I_\gamma$ (rel.)} &
\multicolumn{1}{l}{$R_{\rm DCO}$} &
\multicolumn{1}{l}{$P$} &
\multicolumn{1}{l}{Mult.} &
\multicolumn{1}{l}{$E_i$ (keV)}&
\multicolumn{1}{r}{Band$_i$}&
\multicolumn{1}{l}{$\rightarrow	$}&
\multicolumn{1}{l}{Band$_f$}&
\multicolumn{1}{r}{$I_i^\pi$}&
\multicolumn{1}{l}{$\rightarrow	$}&
\multicolumn{1}{l}{$I_f^\pi$}\\
\hline
\endhead

\hline\hline
\endfoot

\hline
\endlastfoot
234.3(7)  	&	  0.2(1) 	&	                	&	               	&	             	&	6307	&	N4	&$	\rightarrow	$&	N2	&$	35/2^{-}	$&$	\rightarrow	$&$	35/2^{-}	$	\\
251.7(6)  	&	   0.8(1)  	&	                	&	               	&	             	&	5205	&	N4	&$	\rightarrow	$&	N2	&$	31/2^{-}	$&$	\rightarrow	$&$	31/2^{-}	$	\\
290.5(9)  	&	   0.3(1)  	&	               	&	                	&	        	&	5288	&	N7	&$	\rightarrow	$&	N7	&$	29/2^{-}	$&$	\rightarrow	$&$	27/2^{-}	$	\\
326.0(9)  	&	   0.9(2)  	&	               	&	                	&	        	&	5615	&	N7	&$	\rightarrow	$&	N7	&$	31/2^{-}	$&$	\rightarrow	$&$	29/2^{-}	$	\\
350.0(7)  	&	  0.4(1)	&	               	&	                	&	        	&	4210	&	N5	&$	\rightarrow	$&	N1	&$	27/2^{-}	$&$	\rightarrow	$&$	25/2^{-}	$	\\
358.4(6)  	&	   1.4(3)  	&	   1.13(21) 	&	    0.8(4)   	&	   $M1$	&	4159	&	N4	&$	\rightarrow	$&	N2	&$	27/2^{-}	$&$	\rightarrow	$&$	27/2^{-}	$	\\
383.5(7)  	&	   1.6(2)  	&	   0.68(13) 	&	  $-1.6(5)$   	&	    $M1$    	&	2345	&	N4	&$	\rightarrow	$&	N3	&$	19/2^{-}	$&$	\rightarrow	$&$	17/2^{-}	$	\\
384.5(7)  	&	   0.2(1)  	&	   0.63(9)  	&	                	&	   $D$    	&	1356	&	N3	&$	\rightarrow	$&	N2	&$	13/2^{-}	$&$	\rightarrow	$&$	15/2^{-}	$	\\
384.7(14) 	&	  1.8(3)  	&	               	&	                	&	         	&	5998	&	N7	&$	\rightarrow	$&	N7	&$	33/2^{-}	$&$	\rightarrow	$&$	31/2^{-}	$	\\
385.0(7)  	&	   0.7(2)  	&	               	&	                	&	         	&	3285	&	N5	&$	\rightarrow	$&	N1	&$	23/2^{-}	$&$	\rightarrow	$&$	21/2^{-}	$	\\
407.0(6)  	&	   1.2(2)  	&	  0.75(15)  	&	                	&	     $D$    	&	1763	&	N4	&$	\rightarrow	$&	N3	&$	15/2^{-}	$&$	\rightarrow	$&$	13/2^{-}	$	\\
409.0(7)  	&	   1.2(12)  	&	               	&	                	&	            	&	3694	&	N3	&$	\rightarrow	$&	N5	&$	25/2^{-}	$&$	\rightarrow	$&$	23/2^{-}	$	\\
409.4(6)  	&	   1.4(2)  	&	   1.05(19) 	&	   0.6(4) 	&	   $M1$	&	4210	&	N5	&$	\rightarrow	$&	N2	&$	27/2^{-}	$&$	\rightarrow	$&$	27/2^{-}	$	\\
417.4(6)  	&	   1.3(1)  	&	   1.21(15) 	&	   0.8(3) 	&	   $M1$	&	5371	&	N6	&$	\rightarrow	$&	N2	&$	31/2^{-}	$&$	\rightarrow	$&$	31/2^{-}	$	\\
425.0(7)  	&	   0.5(1)  	&	                 	&	                	&	              	&	6498	&	N6	&$	\rightarrow	$&	N2	&$	(35/2^{-})	$&$	\rightarrow	$&$	35/2^{-}	$	\\
430.5(7)  	&	   0.7(2)  	&	   0.67(15) 	&	                	&	    $D$    	&	2776	&	N3	&$	\rightarrow	$&	N4	&$	21/2^{-}	$&$	\rightarrow	$&$	19/2^{-}	$	\\
432.0(9)  	&	   2.2(3)  	&	               	&	                 	&	         	&	6431	&	N7	&$	\rightarrow	$&	N7	&$	35/2^{-}	$&$	\rightarrow	$&$	33/2^{-}	$	\\
453.1(6)  	&	   8.0(12) 	&	   1.11(10)	&	   0.44(9)   	&	   $M1$	&	3154	&	N4	&$	\rightarrow	$&	N2	&$	23/2^{-}	$&$	\rightarrow	$&$	23/2^{-}	$	\\
455.5(11) 	&	   2.0(3)  	&	               	&	                	&	       	&	6887	&	N7	&$	\rightarrow	$&	N7	&$	37/2^{-}	$&$	\rightarrow	$&$	35/2^{-}	$	\\
477.0(11) 	&	   2.0(3)  	&	   0.99(10) 	&	   0.49(19) 	&	   $M1$	&	4278	&	N6	&$	\rightarrow	$&	N2	&$	27/2^{-}	$&$	\rightarrow	$&$	27/2^{-}	$	\\
481.1(6)  	&	  245(87) 	&	   1.00(7)  	&	    0.86(12) 	&	    $E2$    	&	970	&	N2	&$	\rightarrow	$&	N2	&$	15/2^{-}	$&$	\rightarrow	$&$	11/2^{-}	$	\\
496.2(9)  	&	   2.4(3)  	&	   0.39(6)  	&	   $-0.51(16)$ 	&	    $M1$    	&	5279	&	N5	&$	\rightarrow	$&	N3	&$	31/2^{-}	$&$	\rightarrow	$&$	29/2^{-}	$	\\
505.8(6)  	&	  0.9(2)  	&	  	        	&	                  	&	                	&	4784	&	N3	&$	\rightarrow	$&	N6	&$	29/2^{-}	$&$	\rightarrow	$&$	27/2^{-}	$	\\
525.0(11) 	&	   2.6(2)  	&	              	&	                 	&	          	&	7412	&	N7	&$	\rightarrow	$&	N7	&$	39/2^{-}	$&$	\rightarrow	$&$	37/2^{-}	$	\\
530.1(7)  	&	    1.3(2)  	&	              	&	                 	&	          	&	2492	&	N6	&$	\rightarrow	$&	N3	&$	19/2^{-}	$&$	\rightarrow	$&$	17/2^{-}	$	\\
540.0(9)  	&	    0.8(1)  	&	   0.74(11) 	&	               	&	    $D$     	&	3694	&	N3	&$	\rightarrow	$&	N4	&$	25/2^{-}	$&$	\rightarrow	$&$	23/2^{-}	$	\\
552.1(8)  	&	    0.3(1)  	&	               	&	               	&	          	&	7744	&	N6	&$	\rightarrow	$&	N2	&$	(39/2^{-})	$&$	\rightarrow	$&$	39/2^{-}	$	\\
569.1(6)  	&	    2.2(5)  	&	               	&	                	&	         	&	5848	&	N3	&$	\rightarrow	$&	N5	&$	33/2^{-}	$&$	\rightarrow	$&$	31/2^{-}	$	\\
573.4(5)  	&	    3.5(7)  	&	   0.48(8)  	&	     $-0.8(4)$   	&	   $M1$    	&	4784	&	N3	&$	\rightarrow	$&	N5	&$	29/2^{-}	$&$	\rightarrow	$&$	27/2^{-}	$	\\
579.0(7)  	&	    1.6(4)  	&	   0.76(19) 	&	                	&	    $D$    	&	6887	&	N7	&$	\rightarrow	$&	N4	&$	37/2^{-}	$&$	\rightarrow	$&$	35/2^{-}	$	\\
581.7(7)  	&	  9.0(13)	&	   1.00(9)  	&	     0.20(15)  	&	   $E2$    	&	2345	&	N4	&$	\rightarrow	$&	N4	&$	19/2^{-}	$&$	\rightarrow	$&$	15/2^{-}	$	\\
582.5(11) 	&	   1.9(4)  	&	               	&	                	&	         	&	6431	&	N7	&$	\rightarrow	$&	N3	&$	35/2^{-}	$&$	\rightarrow	$&$	33/2^{-}	$	\\
584.8(9)  	&	    3.7(6)  	&	    1.0(4)  	&	    0.21(14)	&	   $M1$	&	3285	&	N5	&$	\rightarrow	$&	N2	&$	23/2^{-}	$&$	\rightarrow	$&$	23/2^{-}	$	\\
600.4(5)  	&	    2.0(4)  	&	               	&	                	&	             	&	8012	&	N7	&$	\rightarrow	$&	N7	&$	41/2^{-}	$&$	\rightarrow	$&$	39/2^{-}	$	\\
603.0(6)  	&	  22.0(18)	&	    1.05(8)  	&	    0.38(13) 	&	  $M1$	&	2345	&	N4	&$	\rightarrow	$&	N2	&$	19/2^{-}	$&$	\rightarrow	$&$	19/2^{-}	$	\\
604.0(5)  	&	    1.4(2)  	&	               	&	                	&	             	&	8616	&	N7	&$	\rightarrow	$&	N7	&$	43/2^{-}	$&$	\rightarrow	$&$	41/2^{-}	$	\\
605.0(7)  	&	    2.5(4)  	&	   1.09(13) 	&	   0.8(3) 	&	   $E2$    	&	1962	&	N3	&$	\rightarrow	$&	N3	&$	17/2^{-}	$&$	\rightarrow	$&$	13/2^{-}	$	\\
616.5(5)  	&	    0.4(1)  	&	               	&	               	&	                 	&	5615	&	N7	&$	\rightarrow	$&	N7	&$	31/2^{-}	$&$	\rightarrow	$&$	27/2^{-}	$	\\
624.8(7)	&	0.5(2)	&		&		&		&	4784	&	N3	&$	\rightarrow	$&	N4	&$	29/2^{-}	$&$	\rightarrow	$&$	27/2^{-}	$	\\
632.0(7)  	&	    4.7(4)  	&	   1.02(8)  	&	   0.69(24) 	&	   $E2$    	&	3285	&	N5	&$	\rightarrow	$&	N5	&$	23/2^{-}	$&$	\rightarrow	$&$	19/2^{-}	$	\\
633.0(7)  	&	    2.7(3)  	&	   1.03(13) 	&	   0.7(3) 	&	  $M1$ 	&	3333	&	N6	&$	\rightarrow	$&	N2	&$	23/2^{-}	$&$	\rightarrow	$&$	23/2^{-}	$	\\
643.1(5)  	&	   0.6(2)   	&	                	&	               	&	          	&	5848	&	N3	&$	\rightarrow	$&	N4	&$	33/2^{-}	$&$	\rightarrow	$&$	31/2^{-}	$	\\
680.3(5)  	&	    0.8(1)  	&	   1.00(12) 	&	               	&	   $E2$      	&	3333	&	N6	&$	\rightarrow	$&	N5	&$	23/2^{-}	$&$	\rightarrow	$&$	19/2^{-}	$	\\
685.1(5)  	&	    0.5(1)  	&	               	&	               	&	            	&	9997	&	N7	&$	\rightarrow	$&	N7	&$	47/2^{-}	$&$	\rightarrow	$&$	45/2^{-}	$	\\
688.5(7)  	&	    1.5(2)  	&	               	&	               	&	              	&	6996	&	N3	&$	\rightarrow	$&	N4	&$	37/2^{-}	$&$	\rightarrow	$&$	35/2^{-}	$	\\
695.9(5)  	&	    0.7(1)  	&	               	&	               	&	         	&	9312	&	N7	&$	\rightarrow	$&	N7	&$	45/2^{-}	$&$	\rightarrow	$&$	43/2^{-}	$	\\
711.0(7)  	&	   0.8(2) 	&	               	&	               	&	         	&	5998	&	N7	&$	\rightarrow	$&	N7	&$	33/2^{-}	$&$	\rightarrow	$&$	29/2^{-}	$	\\
719.5(7)  	&	    1.3(4)  	&	               	&	               	&	         	&	5998	&	N7	&$	\rightarrow	$&	N5	&$	33/2^{-}	$&$	\rightarrow	$&$	31/2^{-}	$	\\
771.7(7)  	&	   180(45) 	&	   0.98(7)  	&	    0.64 (7)	&	   $E2$    	&	1742	&	N2	&$	\rightarrow	$&	N2	&$	19/2^{-}	$&$	\rightarrow	$&$	15/2^{-}	$	\\
779.4(5)  	&	    0.3(1)  	&	               	&	               	&	         	&	10777	&	N7	&$	\rightarrow	$&	N7	&$	49/2^{-}	$&$	\rightarrow	$&$	47/2^{-}	$	\\
793.0(8)  	&	    5.4(19) 	&	              	&	              	&	            	&	1763	&	N4	&$	\rightarrow	$&	N2	&$	15/2^{-}	$&$	\rightarrow	$&$	15/2^{-}	$	\\
793.0(7)  	&	    3.5(3)  	&	   1.09(11) 	&	    0.91(20)	&	   $E2$    	&	3285	&	N5	&$	\rightarrow	$&	N6	&$	23/2^{-}	$&$	\rightarrow	$&$	19/2^{-}	$	\\
793.2(6)  	&	    1.6(2)  	&	   1.09(16) 	&	    0.9(3)	&	   $E2$    	&	3694	&	N3	&$	\rightarrow	$&	N1	&$	25/2^{-}	$&$	\rightarrow	$&$	21/2^{-}	$	\\
793.5(16) 	&	    0.2(1)  	&	   1.04(21) 	&	               	&	   $E2$    	&	4953	&	N2	&$	\rightarrow	$&	N4	&$	31/2^{-}	$&$	\rightarrow	$&$	27/2^{-}	$	\\
794.5(16) 	&	    0.7(1)  	&	                	&	              	&	         	&	5998	&	N7	&$	\rightarrow	$&	N4	&$	33/2^{-}	$&$	\rightarrow	$&$	31/2^{-}	$	\\
808.7(6)  	&	   15.0(15)	&	   0.97(9)  	&	    0.98(14)	&	   $E2$    	&	3154	&	N4	&$	\rightarrow	$&	N4	&$	23/2^{-}	$&$	\rightarrow	$&$	19/2^{-}	$	\\
814.5(7)  	&	    9.0(1) 	&	   1.03(10) 	&	    0.91(21)	&	   $E2$    	&	2776	&	N3	&$	\rightarrow	$&	N3	&$	21/2^{-}	$&$	\rightarrow	$&$	17/2^{-}	$	\\
816.3(5)  	&	    1.8(3)  	&	                	&	               	&	          	&	6431	&	N7	&$	\rightarrow	$&	N7	&$	35/2^{-}	$&$	\rightarrow	$&$	31/2^{-}	$	\\
830.7(9)  	&	    1.5(4)  	&	   0.57(17) 	&	                	&	    $D$    	&	5615	&	N7	&$	\rightarrow	$&	N3	&$	31/2^{-}	$&$	\rightarrow	$&$	29/2^{-}	$	\\
841.7(5)  	&	    3.6(4)  	&	   1.13(11) 	&	    0.7(3) 	&	   $E2$    	&	3333	&	N6	&$	\rightarrow	$&	N6	&$	23/2^{-}	$&$	\rightarrow	$&$	19/2^{-}	$	\\
849.9(5)  	&	     0.8(2) 	&	               	&	                   	&	        	&	5848	&	N3	&$	\rightarrow	$&	N7	&$	33/2^{-}	$&$	\rightarrow	$&$	27/2^{-}	$	\\
867.1(8)  	&	    1.4(6)  	&	               	&	                 	&	          	&	1356	&	N3	&$	\rightarrow	$&	N2	&$	13/2^{-}	$&$	\rightarrow	$&$	11/2^{-}	$	\\
867.5(7)  	&	    2.0(4)  	&	   1.05(17) 	&	                	&	   $E2$    	&	6073	&	N2	&$	\rightarrow	$&	N4	&$	35/2^{-}	$&$	\rightarrow	$&$	31/2^{-}	$	\\
874.9(21) 	&	   1.7(3)  	&	                	&	                	&	          	&	4159	&	N4	&$	\rightarrow	$&	N5	&$	27/2^{-}	$&$	\rightarrow	$&$	23/2^{-}	$	\\
876.3(7)  	&	    4.5(4)  	&	   1.13(12) 	&	    0.71(24) 	&	   $E2$    	&	4210	&	N5	&$	\rightarrow	$&	N6	&$	27/2^{-}	$&$	\rightarrow	$&$	23/2^{-}	$	\\
889.5(11) 	&	   1.4(8)  	&	   1.03(24) 	&	    0.69(21) 	&	   $E2$    	&	2653	&	N5	&$	\rightarrow	$&	N4	&$	19/2^{-}	$&$	\rightarrow	$&$	15/2^{-}	$	\\
889.8(7)  	&	    1.3(4)  	&	               	&	                 	&	          	&	6887	&	N7	&$	\rightarrow	$&	N7	&$	37/2^{-}	$&$	\rightarrow	$&$	33/2^{-}	$	\\
918.4(6)  	&	  12.5(19)	&	    0.96(8)  	&	    0.64(13) 	&	   $E2$    	&	3694	&	N3	&$	\rightarrow	$&	N3	&$	25/2^{-}	$&$	\rightarrow	$&$	21/2^{-}	$	\\
923.7(9)  	&	    2.7(4)  	&	   1.00(7)  	&	                 	&	   $E2$    	&	4784	&	N3	&$	\rightarrow	$&	N1	&$	29/2^{-}	$&$	\rightarrow	$&$	25/2^{-}	$	\\
925.0(6)  	&	    4.8(6)  	&	   0.97(18) 	&	    1.2(3) 	&	   $E2$    	&	4210	&	N5	&$	\rightarrow	$&	N5	&$	27/2^{-}	$&$	\rightarrow	$&$	23/2^{-}	$	\\
927.0(9)  	&	    3.8(6)  	&	   1.06(16) 	&	    1.2(3) 	&	   $E2$   	&	5205	&	N4	&$	\rightarrow	$&	N6	&$	31/2^{-}	$&$	\rightarrow	$&$	27/2^{-}	$	\\
939.2(9)  	&	    5.8(4)  	&	   0.99(14) 	&	    0.8(3) 	&	   $E2$    	&	2901	&	N1	&$	\rightarrow	$&	N3	&$	21/2^{-}	$&$	\rightarrow	$&$	17/2^{-}	$	\\
940.1(5)    	&	     2.5(2)   	&	                  	&	               	&	           	&	3285	&	N5	&$	\rightarrow	$&	N4	&$	23/2^{-}	$&$	\rightarrow	$&$	19/2^{-}	$	\\
945.0(5)  	&	    6.3(6)  	&	   1.07(16) 	&	    0.90(19) 	&	   $E2$    	&	4278	&	N6	&$	\rightarrow	$&	N6	&$	27/2^{-}	$&$	\rightarrow	$&$	23/2^{-}	$	\\
958.4(6)  	&	   145(18) 	&	   1.01(7)  	&	    0.64(7)  	&	   $E2$    	&	2700	&	N2	&$	\rightarrow	$&	N2	&$	23/2^{-}	$&$	\rightarrow	$&$	19/2^{-}	$	\\
959.0(21) 	&	    3.3(3)  	&	               	&	                	&	         	&	3860	&	N1	&$	\rightarrow	$&	N1	&$	25/2^{-}	$&$	\rightarrow	$&$	21/2^{-}	$	\\
980.0(30) 	&	   0.4(1) 	&	   1.19(15) 	&	                	&	    $D+Q$    	&	4784	&	N3	&$	\rightarrow	$&	N2	&$	29/2^{-}	$&$	\rightarrow	$&$	27/2^{-}	$	\\
980.7(6)  	&	    1.5(3)  	&	               	&	                	&	         	&	7412	&	N7	&$	\rightarrow	$&	N7	&$	39/2^{-}	$&$	\rightarrow	$&$	35/2^{-}	$	\\
991.4(6)  	&	   18.0(11)	&	  1.02(12)      	&	   $-0.35(40)$    	&	  $M1+E2$  	&	1962	&	N3	&$	\rightarrow	$&	N2	&$	17/2^{-}	$&$	\rightarrow	$&$	15/2^{-}	$	\\
993.3(5)   	&	       1.5(2)	&	                  	&	                          	&	          	&	4278	&	N6	&$	\rightarrow	$&	N5	&$	27/2^{-}	$&$	\rightarrow	$&$	23/2^{-}	$	\\
993.5(7)  	&	    5.7(9)  	&	 0.85(8)       	&	   $-0.17(22)$    	&	  $M1+E2$  	&	3694	&	N3	&$	\rightarrow	$&	N2	&$	25/2^{-}	$&$	\rightarrow	$&$	23/2^{-}	$	\\
995.3(11) 	&	   4.0(6)  	&	                	&	                	&	         	&	5205	&	N4	&$	\rightarrow	$&	N5	&$	31/2^{-}	$&$	\rightarrow	$&$	27/2^{-}	$	\\
1001.2(14)	&	   2.5(4)  	&	   0.97(18) 	&	    0.70(14) 	&	   $E2$    	&	5279	&	N5	&$	\rightarrow	$&	N6	&$	31/2^{-}	$&$	\rightarrow	$&$	27/2^{-}	$	\\
1005.4(6) 	&	  10.8(10)	&	   0.98(10) 	&	    0.65(17) 	&	   $E2$    	&	4159	&	N4	&$	\rightarrow	$&	N4	&$	27/2^{-}	$&$	\rightarrow	$&$	23/2^{-}	$	\\
1027.5(16)	&	   2.7(4)  	&	               	&	                	&	         	&	6307	&	N4	&$	\rightarrow	$&	N5	&$	35/2^{-}	$&$	\rightarrow	$&$	31/2^{-}	$	\\
1033.8(6) 	&	  22.0(26)	&	   0.89(6)      	&	  $-0.29(10)$     	&	  $M1+E2$  	&	2776	&	N3	&$	\rightarrow	$&	N2	&$	21/2^{-}	$&$	\rightarrow	$&$	19/2^{-}	$	\\
1039.0(7) 	&	    0.9(1)  	&	              	&	                 	&	         	&	6887	&	N7	&$	\rightarrow	$&	N3	&$	37/2^{-}	$&$	\rightarrow	$&$	33/2^{-}	$	\\
1043.0(16)	&	   0.5(1)  	&	              	&	                 	&	         	&	5998	&	N7	&$	\rightarrow	$&	N1	&$	33/2^{-}	$&$	\rightarrow	$&$	29/2^{-}	$	\\
1045.5(7) 	&	    4.2(8)  	&	   0.97(19) 	&	    0.99(21)	&	   $E2$    	&	5205	&	N4	&$	\rightarrow	$&	N4	&$	31/2^{-}	$&$	\rightarrow	$&$	27/2^{-}	$	\\
1056.0(9) 	&	    2.5(4)  	&	   1.01(25) 	&	    0.52(19)	&	   $E2$    	&	4210	&	N5	&$	\rightarrow	$&	N4	&$	27/2^{-}	$&$	\rightarrow	$&$	23/2^{-}	$	\\
1064.2(6) 	&	   7.0(17) 	&	   1.06(11) 	&	    0.65(24)	&	   $E2$    	&	5848	&	N3	&$	\rightarrow	$&	N3	&$	33/2^{-}	$&$	\rightarrow	$&$	29/2^{-}	$	\\
1069.0(11)   	&	    7.0(15) 	&	   1.00(25) 	&	    0.8(4)	&	   $E2$    	&	5279	&	N5	&$	\rightarrow	$&	N5	&$	31/2^{-}	$&$	\rightarrow	$&$	27/2^{-}	$	\\
1074.0(11)	&	   0.9(6)  	&	   0.96(23) 	&	                	&	   $E2$     	&	6353	&	N5	&$	\rightarrow	$&	N5	&$	35/2^{-}	$&$	\rightarrow	$&$	31/2^{-}	$	\\
1078.0(21)	&	  1.1(5)  	&	               	&	                	&	         	&	5288	&	N7	&$	\rightarrow	$&	N5	&$	29/2^{-}	$&$	\rightarrow	$&$	27/2^{-}	$	\\
1083.0(11)	&	   3.0(5)  	&	   0.98(14) 	&	    1.0(5)	&	   $E2$    	&	3860	&	N1	&$	\rightarrow	$&	N3	&$	25/2^{-}	$&$	\rightarrow	$&$	21/2^{-}	$	\\
1089.6(5) 	&	  12.0(14) 	&	   0.99(8)  	&	    0.99(19)  	&	   $E2$    	&	4784	&	N3	&$	\rightarrow	$&	N3	&$	29/2^{-}	$&$	\rightarrow	$&$	25/2^{-}	$	\\
1092.8(7) 	&	    1.3(2)  	&	              	&	                  	&	         	&	5371	&	N6	&$	\rightarrow	$&	N6	&$	31/2^{-}	$&$	\rightarrow	$&$	27/2^{-}	$	\\
1097.0(21)	&	    2.1(4)  	&	              	&	                 	&	         	&	4955	&	N1	&$	\rightarrow	$&	N1	&$	29/2^{-}	$&$	\rightarrow	$&$	25/2^{-}	$	\\
1100.2(6) 	&	   69.7(52)	&	   0.99(3)  	&	    0.68(5)  	&	   $E2$    	&	3801	&	N2	&$	\rightarrow	$&	N2	&$	27/2^{-}	$&$	\rightarrow	$&$	23/2^{-}	$	\\
1103.0(16)	&	    4.2(6)  	&	  0.99(23) 	&	    0.9(5)  	&	  $E2$    	&	6307	&	N4	&$	\rightarrow	$&	N4	&$	35/2^{-}	$&$	\rightarrow	$&$	31/2^{-}	$	\\
1103.5(16)	&	    1.4(5)  	&	               	&	                	&	         	&	7412	&	N7	&$	\rightarrow	$&	N4	&$	39/2^{-}	$&$	\rightarrow	$&$	35/2^{-}	$	\\
1119.0(11)	&	   12.0(11)	&	   1.00(8)  	&	    0.77(10) 	&	   $E2$     	&	7192	&	N2	&$	\rightarrow	$&	N2	&$	39/2^{-}	$&$	\rightarrow	$&$	35/2^{-}	$	\\
1119.5(7) 	&	   18.0(18)	&	   1.00(8)  	&	    0.77(10)  	&	   $E2$    	&	6073	&	N2	&$	\rightarrow	$&	N2	&$	35/2^{-}	$&$	\rightarrow	$&$	31/2^{-}	$	\\
1119.9(5)  	&	     1.2(2)  	&	                   	&	               	&	          	&	5279	&	N5	&$	\rightarrow	$&	N4	&$	31/2^{-}	$&$	\rightarrow	$&$	27/2^{-}	$	\\
1126.0(21)	&	    1.7(4)  	&	               	&	                	&	          	&	8012	&	N7	&$	\rightarrow	$&	N7	&$	41/2^{-}	$&$	\rightarrow	$&$	37/2^{-}	$	\\
1126.5(8) 	&	     2.0(3)  	&	               	&	                	&	         	&	6498	&	N6	&$	\rightarrow	$&	N6	&$	(35/2^{-})	$&$	\rightarrow	$&$	31/2^{-}	$	\\
1146.5(11)	&	    0.9(1)  	&	   1.03(13) 	&	    0.31(17) 	&	   $E2$    	&	6996	&	N3	&$	\rightarrow	$&	N3	&$	37/2^{-}	$&$	\rightarrow	$&$	33/2^{-}	$	\\
1148.0(16)	&	    3.3(6)  	&	   1.1(3)    	&	                	&	   $E2$    	&	6353	&	N5	&$	\rightarrow	$&	N4	&$	35/2^{-}	$&$	\rightarrow	$&$	31/2^{-}	$	\\
1152.4(6) 	&	   36.5(37)	&	   1.04(8)  	&	    0.68(10)  	&	   $E2$    	&	4953	&	N2	&$	\rightarrow	$&	N2	&$	31/2^{-}	$&$	\rightarrow	$&$	27/2^{-}	$	\\
1154.0(16)	&	   5.0(18) 	&	               	&	                 	&	         	&	4955	&	N1	&$	\rightarrow	$&	N2	&$	29/2^{-}	$&$	\rightarrow	$&$	27/2^{-}	$	\\
1159.0(16)	&	   6.0(13) 	&	               	&	                 	&	         	&	2901	&	N1	&$	\rightarrow	$&	N2	&$	21/2^{-}	$&$	\rightarrow	$&$	19/2^{-}	$	\\
1159.7(16)	&	   5.0(13) 	&	               	&	                 	&	          	&	3860	&	N1	&$	\rightarrow	$&	N2	&$	25/2^{-}	$&$	\rightarrow	$&$	23/2^{-}	$	\\
1187.0(11)	&	    1.1(3)  	&	   1.06(15) 	&	    0.9(5)     	&	   $E2$    	&	7494	&	N4	&$	\rightarrow	$&	N4	&$	39/2^{-}	$&$	\rightarrow	$&$	35/2^{-}	$	\\
1201.0(21)	&	    0.2(1)  	&	               	&	                	&	         	&	8695	&	N4	&$	\rightarrow	$&	N4	&$	(43/2^{-})	$&$	\rightarrow	$&$	39/2^{-}	$	\\
1204.6(6) 	&	    1.7(3)  	&	               	&	                 	&	         	&	8616	&	N7	&$	\rightarrow	$&	N7	&$	43/2^{-}	$&$	\rightarrow	$&$	39/2^{-}	$	\\
1211.3(6) 	&	    0.5(5)  	&	   0.97(20) 	&	                	&	   $E2$    	&	5371	&	N6	&$	\rightarrow	$&	N4	&$	31/2^{-}	$&$	\rightarrow	$&$	27/2^{-}	$	\\
1214.7(6) 	&	    6.8(7)  	&	   0.96(7)  	&	    1.02(18) 	&	   $E2$    	&	8406	&	N2	&$	\rightarrow	$&	N2	&$	43/2^{-}	$&$	\rightarrow	$&$	39/2^{-}	$	\\
1246.2(9) 	&	    0.6(3)  	&	               	&	                 	&	         	&	7744	&	N6	&$	\rightarrow	$&	N6	&$	(39/2^{-})	$&$	\rightarrow	$&$	(35/2^{-})	$	\\
1260.8(7) 	&	    1.8(3)  	&	   1.05(20) 	&	                	&	   $E2$    	&	7614	&	N5	&$	\rightarrow	$&	N5	&$	39/2^{-}	$&$	\rightarrow	$&$	35/2^{-}	$	\\
1261.5(16)	&	   2.5(4)  	&	   0.98(17) 	&	    1.1(5)  	&	   $E2$    	&	4955	&	N1	&$	\rightarrow	$&	N3	&$	29/2^{-}	$&$	\rightarrow	$&$	25/2^{-}	$	\\
1273.8(13)	&	  4.2(15) 	&	                	&	                 	&	         	&	1763	&	N4	&$	\rightarrow	$&	N2	&$	15/2^{-}	$&$	\rightarrow	$&$	11/2^{-}	$	\\
1300.1(5) 	&	   1.0(2)  	&	                	&	                 	&	         	&	9312	&	N7	&$	\rightarrow	$&	N7	&$	45/2^{-}	$&$	\rightarrow	$&$	41/2^{-}	$	\\
1302.0(11)	&	   0.5(1)  	&	               	&	                 	&	         	&	8298	&	N3	&$	\rightarrow	$&	N3	&$	(41/2^{-})	$&$	\rightarrow	$&$	37/2^{-}	$	\\
1305.0(6) 	&	   0.4(1)  	&	                	&	                 	&	         	&	4998	&	N7	&$	\rightarrow	$&	N3	&$	27/2^{-}	$&$	\rightarrow	$&$	25/2^{-}	$	\\
1317.8(7) 	&	   3.2(3)  	&	   0.96(9)   	&	    0.76(17)  	&	   $E2$    	&	9724	&	N2	&$	\rightarrow	$&	N2	&$	47/2^{-}	$&$	\rightarrow	$&$	43/2^{-}	$	\\
1336.0(16)	&	   0.7(1)  	&	               	&	                 	&	         	&	8950	&	N5	&$	\rightarrow	$&	N5	&$	(43/2^{-})	$&$	\rightarrow	$&$	39/2^{-}	$	\\
1354.0(7) 	&	   1.6(2)  	&	               	&	                 	&	          	&	6307	&	N4	&$	\rightarrow	$&	N2	&$	35/2^{-}	$&$	\rightarrow	$&$	31/2^{-}	$	\\
1370.0(21)	&	  0.2(1) 	&	               	&	                 	&	         	&	10066	&	N4	&$	\rightarrow	$&	N4	&$	(47/2^{-})	$&$	\rightarrow	$&$	(43/2^{-})	$	\\
1374.3(8) 	&	   2.0(3)  	&	   1.06(17) 	&	                  	&	   $E2$    	&	2345	&	N4	&$	\rightarrow	$&	N2	&$	19/2^{-}	$&$	\rightarrow	$&$	15/2^{-}	$	\\
1380.7(9) 	&	   1.7(2)  	&	   1.03(12) 	&	    0.7(3)    	&	   $E2$    	&	11105	&	N2	&$	\rightarrow	$&	N2	&$	51/2^{-}	$&$	\rightarrow	$&$	47/2^{-}	$	\\
1381.7(6) 	&	   1.0(2)  	&	               	&	                  	&	         	&	9997	&	N7	&$	\rightarrow	$&	N7	&$	47/2^{-}	$&$	\rightarrow	$&$	43/2^{-}	$	\\
1400.5(11)	&	   2.5(3)  	&	   1.08(22) 	&	    0.6(4) 	&	   $E2$    	&	6353	&	N5	&$	\rightarrow	$&	N2	&$	35/2^{-}	$&$	\rightarrow	$&$	31/2^{-}	$	\\
1406.3(21)	&	   0.7(2)  	&	               	&	                 	&	         	&	5205	&	N4	&$	\rightarrow	$&	N2	&$	31/2^{-}	$&$	\rightarrow	$&$	27/2^{-}	$	\\
1412.0(21)	&	   1.8(2)  	&	   0.95(16)	&	                  	&	   $E2$     	&	3154	&	N4	&$	\rightarrow	$&	N2	&$	23/2^{-}	$&$	\rightarrow	$&$	19/2^{-}	$	\\
1424.0(21)	&	   0.8(2)  	&	               	&	                 	&	         	&	6379	&	N1	&$	\rightarrow	$&	N1	&$	(33/2^{-})	$&$	\rightarrow	$&$	29/2^{-}	$	\\
1458.5(11)	&	   2.2(6)  	&	              	&	                  	&	         	&	4159	&	N4	&$	\rightarrow	$&	N2	&$	27/2^{-}	$&$	\rightarrow	$&$	23/2^{-}	$	\\
1461.0(11)	&	   0.5(1)  	&	              	&	                  	&	         	&	12566	&	N2	&$	\rightarrow	$&	N2	&$	(55/2^{-})	$&$	\rightarrow	$&$	51/2^{-}	$	\\
1463.4(15)	&	   0.5(1)  	&	              	&	                  	&	         	&	10777	&	N7	&$	\rightarrow	$&	N7	&$	49/2^{-}	$&$	\rightarrow	$&$	45/2^{-}	$	\\
1477.0(21)	&	   1.0(4)  	&	   0.96(23) 	&	                 	&	   $E2$    	&	6431	&	N7	&$	\rightarrow	$&	N2	&$	35/2^{-}	$&$	\rightarrow	$&$	31/2^{-}	$	\\
1478.5(11)	&	   2.5(5)  	&	   1.05(22) 	&	                 	&	   $E2$    	&	5279	&	N5	&$	\rightarrow	$&	N2	&$	31/2^{-}	$&$	\rightarrow	$&$	27/2^{-}	$	\\
1486.0(21)	&	   0.3(2)  	&	               	&	                 	&	         	&	5288	&	N7	&$	\rightarrow	$&	N2	&$	29/2^{-}	$&$	\rightarrow	$&$	27/2^{-}	$	\\
1509.8(13)	&	  9.0(10) 	&	   0.98(15) 	&	    0.74(20)  	&	   $E2$    	&	4210	&	N5	&$	\rightarrow	$&	N2	&$	27/2^{-}	$&$	\rightarrow	$&$	23/2^{-}	$	\\
1521.3(6) 	&	  11.0(26)	&	   0.94(10) 	&	    0.69(20)  	&	   $E2$    	&	2492	&	N6	&$	\rightarrow	$&	N2	&$	19/2^{-}	$&$	\rightarrow	$&$	15/2^{-}	$	\\
1541.5(11)	&	  0.2(1) 	&	               	&	                  	&	         	&	14107	&	N2	&$	\rightarrow	$&	N2	&$	(59/2^{-})	$&$	\rightarrow	$&$	(55/2^{-})	$	\\
1569.5(11)	&	   1.5(9)  	&	   0.99(22) 	&	                 	&	   $E2$    	&	5371	&	N6	&$	\rightarrow	$&	N2	&$	31/2^{-}	$&$	\rightarrow	$&$	27/2^{-}	$	\\
1578.0(7) 	&	    5.5(6)  	&	   1.11(20) 	&	    1.3(4)    	&	   $E2$    	&	4278	&	N6	&$	\rightarrow	$&	N2	&$	27/2^{-}	$&$	\rightarrow	$&$	23/2^{-}	$	\\
1591.5(7) 	&	  11.5(11)	&	   1.08(10) 	&	    1.2(3)     	&	   $E2$    	&	3333	&	N6	&$	\rightarrow	$&	N2	&$	23/2^{-}	$&$	\rightarrow	$&$	19/2^{-}	$	\\
1618.0(21)	&	  0.1(1) 	&	                	&	                 	&	         	&	15726	&	N2	&$	\rightarrow	$&	N2	&$	(63/2^{-})	$&$	\rightarrow	$&$	(59/2^{-})	$	\\
1683.0(9) 	&	   4.1(10) 	&	   1.00(12) 	&	    1.0(3)  	&	   $E2$    	&	2653	&	N5	&$	\rightarrow	$&	N2	&$	19/2^{-}	$&$	\rightarrow	$&$	15/2^{-}	$	\\
1815.0(21)	&	  0.2(1) 	&	    1.1(3)   	&	                	&	    $E2$     	&	5615	&	N7	&$	\rightarrow	$&	N2	&$	31/2^{-}	$&$	\rightarrow	$&$	27/2^{-}	$	\\

\hline
\hline
\end{longtable*}
\vskip -0.5cm
\noindent {\footnotesize $D$ indicates stretched dipole transitions.}
\\ {\footnotesize $D+Q$ indicates mixed dipole + quadrupole transitions.}
\\

\section{Negative-parity level scheme}

Medium- and high-spin states of $^{105}$Pd have previously 
been studied by Rickey~\textit{et al.}~\cite{105Pd-1} using 
the asymmetric $^{96}\textrm{Zr}(^{12}\textrm{C}, 3n)$ fusion evaporation 
reaction, while Macchiavelli~\textit{et al.}~\cite{105Pd-2}
have studied the high-spin superdeformed bands of this nucleus 
using the rather symmetric $^{64}\textrm{Ni}(^{48}\textrm{Ca}, {\alpha}3n)$
reaction. In the present work, we used the $^{96}\textrm{Zr}(^{13}\textrm{C}, 4n)$ 
reaction, which is very similar to the one used by Rickey~\textit{et al.}, 
thus we also excited the medium- and high-spin states, 
but could not excite the superdeformed bands. However, 
the present high-statistics experiment enabled us to 
significantly extend the medium- and high-spin level 
scheme of $^{105}$Pd.

\begin{figure*}
\centering
\includegraphics[width=15.0cm, bb=130 0 660 600]{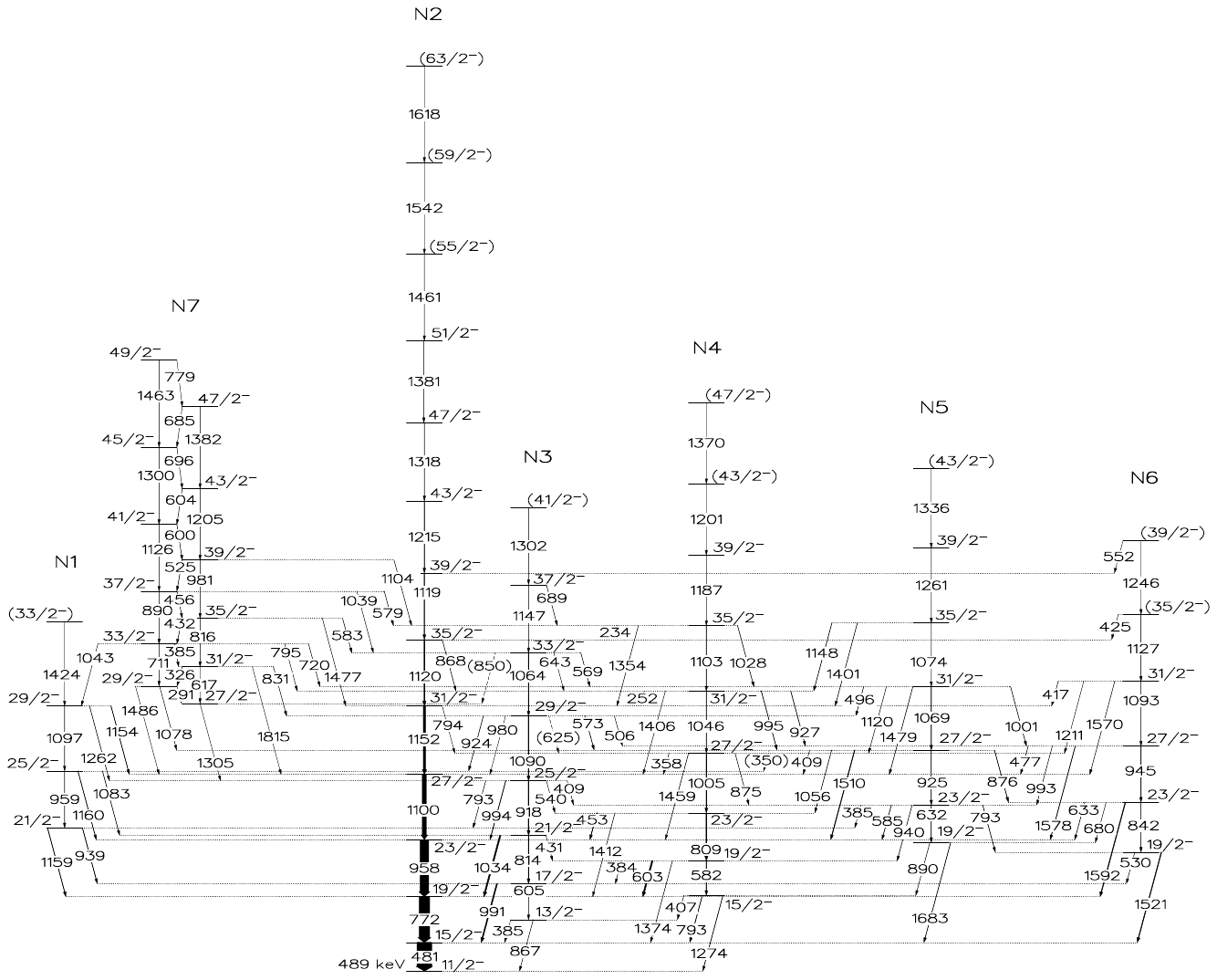}
\vspace{-1.5cm}
\caption{The partial level scheme of $^{105}$Pd obtained in the present work.
Widths of the lines are proportional with the transition intensities.}
\label{ls}
\end{figure*}

The proposed negative-parity level scheme is shown in Fig.~\ref{ls}. It was constructed using the {$E_{\gamma}$}-{$E_{\gamma}$}-{$E_{\gamma}$} coincidence relations, as well as energy and intensity balances. When no other information was available, the order of the transitions in the $\gamma$-ray cascade was deduced based on their intensities. The placement of the known transitions is consistent with the previous work of Rickey~\textit{et al.} except that our coincidence data did not support the existence of the 183-, 962-, 749- and 1311-keV $\gamma$ rays. In Ref.~\cite{105Pd} we previously published bands N1, N2, and N3, and now we have extended them to higher excitation energies and spins. We have also reported previously the observation of band N4 in Ref.~\cite{105Pd-cp}.
Furthermore, new bands of N5, N6 and N7 have been observed and linked to the previously reported ones.

Band N2 was previously reported up to $I^{\pi}= 43/2^{-}$ in Refs.~\cite{105Pd-1,105Pd-2}. Our experimental data confirm the previous placements of the transitions assigned to this band and the spin-parity values of the levels up to $I^{\pi} = 43/2^{-}$. We have extended the band by five more levels. From the DCO and linear polarization data, stretched $E2$ characters were deduced for the 1318- and 1381-keV $\gamma$ rays resulting in unambiguous $I^{\pi}=47/2^{-}$ and $51/2^{-}$ spin-parities for the states at 9724 and 11105~keV. The 1461-, 1542-, and 1618-keV $\gamma$ rays are too weak to determine their multipolarities. However, as these $\gamma$ rays continue the rotational sequence, they are assumed to be stretched $E2$ transitions and tentative spin-parity values of
$I^{\pi} = (55/2^{-})$, $(59/2^{-})$, and $(63/2^{-})$ are assigned to the states at 12566, 14107, and 15726-keV excitation energies.

Only the first state of band N1 with $I^{\pi} = 21/2^{-}$ was reported by 
Rickey \textit{et al.}~\cite{105Pd-1} earlier. Previously, we extended this band with two additional levels up to $I^{\pi}=29/2^{-}$ spin-parity and $E_{x}= 4955$-keV excitation energy~\cite{105Pd}. In the present work, we have added one more level to the top of band N1 decaying by the weak 1424-keV $\gamma$ ray. We have performed DCO and linear polarization analysis for the 939-, 1083-, and 1262-keV transitions connecting band N1 to band N2. Stretched $E2$ characters have been obtained for these $\gamma$ rays supporting the previous spin-parity assignments of the states at 2901, 3860, and 4955~keV. Due to the lack of statistics no information could be obtained on the multipolarity of the new 1424-keV transition. Since the rotational sequence continues, we suggest $I^{\pi} = (33/2^{-})$ spin-parity value for the state at $E_{x} = 6379$-keV excitation energy.

Band N3 was observed up to $I^{\pi} = 25/2^{-}$ in Ref.~\cite{105Pd-1}. In our previous publication, we reported the extension of this band to 41/2$^{-}$ spin-parity value~\cite{105Pd}. In the present work, we have identified nine additional $\gamma$ rays with the energy of 409, 431, 506, 540, 569, 573, 625, 643, and 689~keV to band N3, which connect it to the new bands, namely to bands N4 and N5. DCO ratios and linear polarization values have been deduced for several already known transitions in band N3. Based on the obtained results we determined unambiguous spin-parity until $I^{\pi} = 37/2^{-}$ to the states in band N3 consistent with previous assignments. Due to the lack of statistics tentative $I^{\pi} = (41/2^{-})$ is suggested to the state at 8298~keV. Among the $\gamma$ rays identified to band N3 in the present work, stretched $M1$ multipolarity has been determined for the 573-keV transition and stretched dipole character has been obtained for the 431- and 540-keV transitions by DCO and linear polarization analysis. This information has been used in the spin-parity assigments of the states in bands N4 and N5. The intensity of the 409-, 569-, and 689-keV $\gamma$ rays is not sufficient to draw conclusions about their multipolarity.

The three lowest-lying states in band N4 up to spin-parity $I^{\pi} = 23/2^{-}$ were reported by Rickey \textit{et al.}~\cite{105Pd-1}. Our coincidence analysis, DCO and linear polarization results strengthen the placement of the known transitions and the spin-parity assignment of the already identified excited sates. In addition, we have significantly extended this band establishing six new levels. For the newly observed 1005-, 1046-, 1103-, and 1187-keV transitions stretched $E2$ character has been deduced by DCO and linear polarization analysis. Therefore, we propose $I^{\pi} = 27/2^{-}$, 31/2$^{-}$, $35/2^{-}$, and $39/2^{-}$ spin-parity values for the states at excitation energies of 4159, 5205, 6307, 7494~keV. Due to the lack of statistics multipolarity could not be determined for the 1201- and 1370-keV $\gamma$ rays. However, as these transitions continue the rotational structure, we suggest tentative $I^{\pi} = (43/2^{-})$ and ($47/2^{-}$) spin-parity values for the levels established by them. Furthermore, we have observed new transitions connecting band N4 to bands N1, N2, N3, N5, and N6. We have determined DCO ratios and linear polarization values for some of these transitions (see in Table~\ref{table1}), which are consistent with previous spin-parities and support our new assignments. Karmakar \textit{et al.} reported in the arXiv~\cite{Karmref} on the observation
of two new transitions from this band to band N1 and another new one to
band N3 compared to the previously published transitions in Ref.~\cite{105Pd-cp}. Those new transitions were not
observed in the present work.

Band N5 is a new rotational structure assigned to $^{105}$Pd in the present work. We have established seven levels within band N5 and identified several transitions linking it to the other bands. Our finding indicates that the 890- and 1683-keV $\gamma$ rays decaying from the first state in band N5 to the $I^{\pi} = 15/2^{-}$ states in bands N2 and N4, are stretched $E2$ transitions. This implies a 19/2$^{-}$ spin-parity for the state at 2653~keV. For the next five transitions building the rotation cascade on the top of this state we have determined stretched $E2$ multipolarity. Thus, we propose unambiguous spin-parities up to $I^{\pi} = 39/2^{-}$ in band N5. These assignments are supported by the stretched $E2$ character obtained for some of the transitions linking band N5 to bands N2 and N4. Since the 1336-keV $\gamma$ ray that depopulates the highest-lying state in band N5 was too weak to deduce its multipolarity, we suggest a tentative value of $I^{\pi} = (43/2^{-})$ for this state as the 1336-keV $\gamma$ ray continues the rotational sequence. Band N5 is connected to band N6 by the 793-, 876-, and 1001-keV transitions which are found to have stretched $E2$ character. These assignments have been used in the spin-parity determination of the excited states in band N6.

In band N6 Rickey \textit{et al.}~\cite{105Pd-1} identified the lowest-lying 749- and 1521-keV $\gamma$ rays. On the basis of our coincidence analysis only the 1521-keV transition could be confirmed. In addition, we have built a rotational structure with five levels on the top of the excited state at 2492~keV. This new band is connected to bands N2, N4, and N5 by several transitions. The $I^{\pi} = 19/2^{-}$ spin-parity of the first state in band N6 is fixed by the stretched $E2$ multipolarity of the 1521-keV transition connecting it to the first excited state with 15/2$^{-}$ in band N2. We deduced stretched $E2$ character also for the following two transitions in the rotational sequence with the energy of 842 and 945~keV. Thus, $I^{\pi} = 23/2^{-}$ and $27/2^{-}$ values are proposed to the levels at 3333 and 4278~keV. These assignments are strengthened by the stretched $E2$ multipolarity of the 680-, 1578-, and 1592-keV linking transitions. Although the low statistics for the 1093-keV $\gamma$ ray did not allow to determine its multipolarity, the stretched $E2$ character obtained for the 1211- and 1570-keV transitions, decaying parallel to states with unambiguous spin-parities in bands N2 and N4, fixed the $I^{\pi} = 31/2^{-}$ value for the level at 5371~keV. We propose a ($39/2^{-}$) spin-parity for the last state of band N6, since the rotational sequence continues. The obtained multipolarities of $\gamma$ rays feeding the states of band N6 from levels in bands N4 and N5 with already firmly assigned spin-parities strengthen the assignments described above.

Band N7 was observed the first time in the present work. It is linked via several transitions to all the other negative-parity bands except for band N6 which confirms its identification to $^{105}$Pd. Based on the coincidence data we established twelve excited states in band N7 connected by several in-band and cross-over transitions. As band N7 is populated weakly we could deduce only DCO ratios and not linear polarization information for two linking $\gamma$ rays: the 831- and 579-keV were found to have stretched dipole character. As they feed states with $I = 29/2$ and $I = 35/2$ spins, we assign $I = 31/2$ and $I = 37/2$ spins to their initial levels at 5615 and 6887~keV, respectively. The 1815- and 1039-keV $\gamma$ rays decay parallel to the 831- and 579-keV transitions from the same initial states to levels having spin less by 1 unit than the final states of the 831- and 579-keV transitions. It implies that the 1815- and 1039-keV $\gamma$ rays are stretched $E2$ transitions keeping the negative parity for band N7.  Since we managed to find all the in-band and cross-over transitions between the states in the two branches of band N7, we assign negative parity to the whole band and spins from $I = 27/2$ to $I = 49/2$ to the states. The consistency of the level scheme is confirmed by the stretched $E2$ character of the 1043-, 1104-, and 1477-keV transitions connecting band N7 to states in bands N1, N2, and N4 with unambiguous spin-parities.

\LTcapwidth=\textwidth
\setlength{\tabcolsep}{7.5pt}
\renewcommand{\arraystretch}{1.3}
\begin{table*}[!ht]
\caption{The configurations (both valence and unpaired nucleon)
 as well as the corresponding energies (in MeV) and the deformation parameters 
 $\beta$ and $\gamma$ for the local minima in $^{105}$Pd obtained by 
 the configuration-fixed constrained triaxial RDFT calculations.}
\label{tab:Pd105}
\begin{tabular}{ccccccc}
\hline
State    & $E_{\rm x}$& $(\beta, \gamma)$ & $\rm Valence~configuration$ 
& $\rm Unpaired~configuration$ & $\pi$ & Band \\
   \hline
A   &	$0.00$  &  $(0.19, ~0.0^\circ)$
    & $\pi(1g_{9/2})^{6}(2p_{1/2})^2 \otimes \nu(1g_{7/2})^{5}(2d_{5/2})^4$
    & $\nu(1g_{7/2})^{1}$ & $+$\\
B	&	$0.38$	&  $(0.23, 22.2^\circ)$
    & $\pi(1g_{9/2})^{8} \otimes \nu(1g_{7/2})^{5}(2d_{5/2})^4$
    & $\nu(1g_{7/2})^{1}$ & $+$ \\
C	&	$0.51$	&  $(0.27, 24.9^\circ)$
    & $\pi(1g_{9/2})^{8} \otimes \nu(1g_{7/2})^{6}(2d_{3/2})^2(1h_{11/2})^1$
    & $\nu(1h_{11/2})^{1}$  & $-$ & N1, N2, N3\\
D	&	$0.73$	&  $(0.29, 30.7^\circ)$
    & $\pi(1g_{9/2})^{8} \otimes \nu(1g_{7/2})^{3}(2d_{5/2})^2(2d_{3/2})^2(1h_{11/2})^2$
    & $\nu(1g_{7/2})^1$  & $+$ \\ 
\hline
a	&	$1.94$	&  $(0.25, 20.1^\circ)$
    & $\pi(1g_{9/2})^{8} \otimes \nu(1g_{7/2})^{6}(2d_{5/2})^2(1h_{11/2})^1$
    & $\nu(2d_{5/2})^2(1h_{11/2})^1$  & $-$ & N6\\    
b	&	$3.30$	&  $(0.25, 28.0^\circ)$
    & $\pi(1g_{9/2})^{8} \otimes \nu(1g_{7/2})^{6}(2d_{3/2})^2(1h_{11/2})^1$
    & $\pi(1g_{9/2})^{-2} \otimes \nu(1h_{11/2})^1$  & $-$ & N3-high, N5 \\ 
c	&	$4.33$	&  $(0.24, 32.8^\circ)$
    & $\pi(1g_{9/2})^{8} \otimes \nu(1g_{7/2})^{4}(2d_{5/2})^2(2d_{3/2})^2(1h_{11/2})^1$
    & $\pi(1g_{9/2})^{-2}\otimes \nu (1g_{7/2})^{-2}(1h_{11/2})^1$  & $-$ & N7\\ 
d	&	$5.43$	&  $(0.29, 10.3^\circ)$
    & $\pi(1g_{9/2})^{8} \otimes \nu(1g_{7/2})^{4}(2d_{5/2})^2(1h_{11/2})^3$
    & $\nu(1h_{11/2})^3$  & $-$ & N2-high\\ 
\hline
\end{tabular}
\end{table*}

\section{Discussion}

The main aim of the present study was to explore the most complete 
negative-parity band structure possible using the available 
experimental data set, and to interpret it with calculations 
using constrained triaxial relativistic density functional 
theory (RDFT)~\cite{J.Meng2006PRC, J.Meng2016book, J.Meng2023book}
and the quantum particle rotor model (PRM)~\cite{Bohr1975book, 
1, J.Peng2003PRC, S.Q.Zhang2007PRC, B.Qi2009PLB, Q.B.Chen2018PLB, 
Streck2018PRC, Q.B.Chen2018PRC, Q.B.Chen2019PRC, Q.B.Chen2022EPJA}. 
This work also aimed at searching for chiral bands and for the possible 
two-phonon wobbling band in this nucleus. Three new decoupled bands with 
E2 transitions and one strongly coupled band with $M1+E2$ and 
E2 transitions have been observed. Chiral band pair 
could not be found in this search. All the three observed 
decoupled bands have the same parities and level spins which 
are expected for the two-phonon wobbling band. However, 
none of them have been found to decay to the one-phonon 
wobbling band (band N3) by $M1+E2$ transitions of strong $E2$ 
character, which is the signature expected for the two-phonon wobbling band. 

\subsection{Configurations and deformation parameters}

In order to understand the nature of the observed negative-parity 
band structure in $^{105}$Pd, first adiabatic and 
configuration-fixed constrained RDFT calculations~\cite{J.Meng2006PRC, 
J.Meng2016book, J.Meng2023book} were performed to search for 
the possible configurations and deformations. 

\begin{figure}[!ht]
    \centering
    \includegraphics[width=0.90\linewidth]{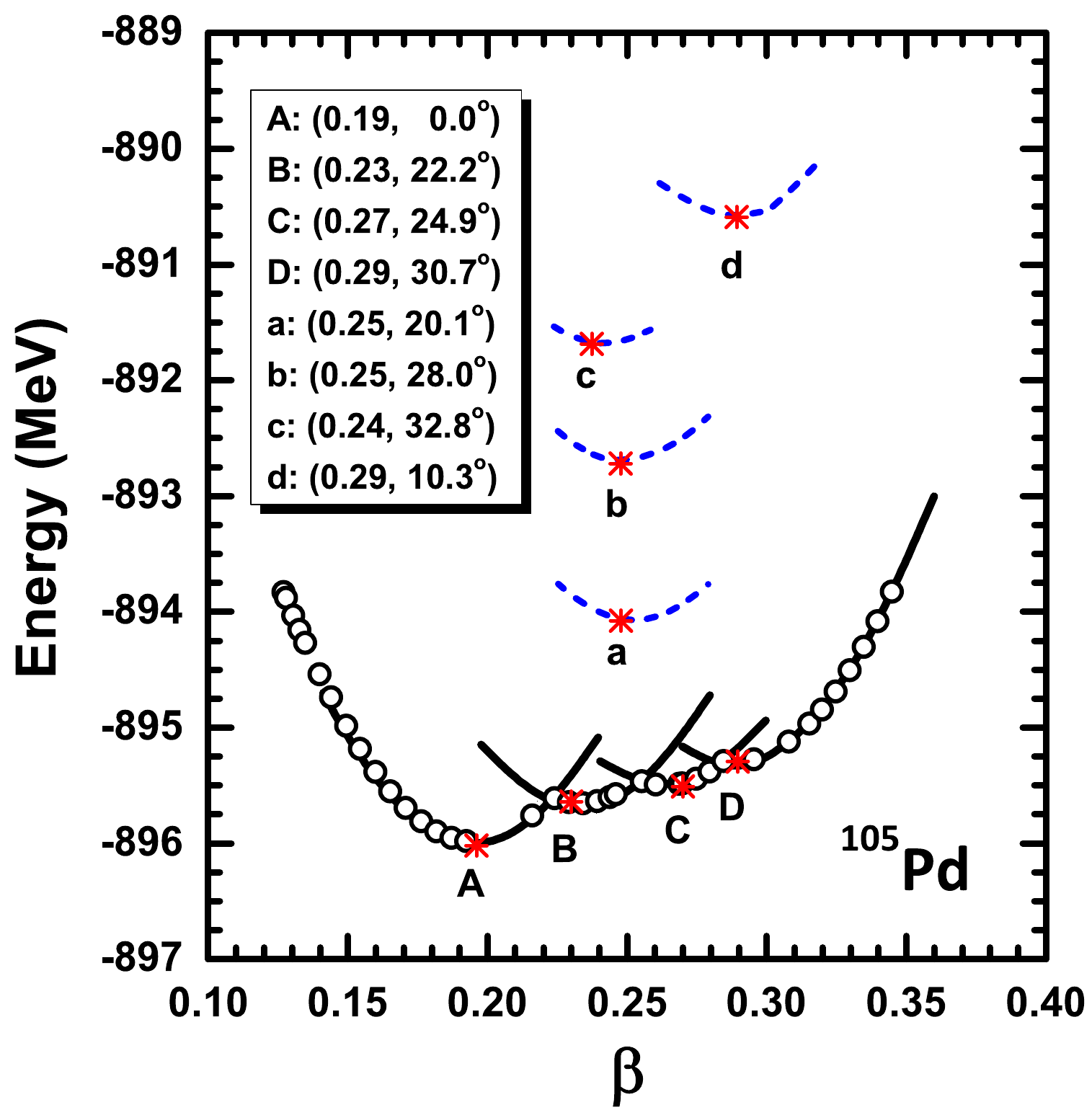}
    \caption{The potential energy curves (PECs) as functions of 
    deformation $\beta$ in adiabatic (circles) and configuration-fixed (lines) 
    constrained triaxial RDFT calculation for $^{105}$Pd. 
    The local minima in the energy surfaces for the fixed 
    configuration are represented as stars. The $\beta_2$ and $\gamma$ deformation parameters are listed for the different configurations in the label.}
    \label{fig:PEC}
\end{figure}

By minimizing the energy with respect to the deformation 
$\gamma$ for a given $\beta$, both adiabatic and 
configuration-fixed constrained triaxial RDFT calculations
have been performed with the effective functional PC-PK1~\cite{P.W.Zhao2010PRC} 
for $^{105}\rm Pd$. The obtained potential energy curves (PECs) 
are shown in Fig.~\ref{fig:PEC}. The adiabatic PEC  
is separated into several regions owing to different configurations 
and the uninterrupted PEC with a certain configuration 
can be derived from the corresponding configuration-fixed
constrained calculations. This gives the local minima A, B, C, 
and D in Fig.~\ref{fig:PEC}. The excitation energy $E_{\rm x}$, 
deformation parameters, valence and unpaired nucleon configurations, 
as well as the parity for these minima are summarized in Table~\ref{tab:Pd105}.
It is found that the minima A-D all correspond to one unpaired 
nucleon configurations. With the exception of  configuration A, the remaining 
configurations possess triaxial deformation. In particular,
the negative parity configuration C was assigned to the low spin 
parts of bands N1, N2, N3 in Ref.~\cite{105Pd} where wobbling motion 
based on an odd-neutron configuration was identified. Furthermore, 
as shown in Fig.~\ref{fig:PEC}, the configuration-fixed constrained 
calculation is also performed for the four multi-unpaired-nucleon
configurations a, b, c, d. Their corresponding information 
is also listed in Table~\ref{tab:Pd105}. These negative parity 
configurations are assigned to the other bands observed
in the present work. 

\subsection{Band N4}

With the obtained configurations and deformations from 
RDFT calculations, the quantal PRM~\cite{Bohr1975book, 
1, J.Peng2003PRC, S.Q.Zhang2007PRC, B.Qi2009PLB, Q.B.Chen2018PLB, 
Streck2018PRC, Q.B.Chen2018PRC, Q.B.Chen2019PRC, Q.B.Chen2022EPJA} 
calculations were performed to study the energy spectra 
and $B(M1)/B(E2)$ ratios.

\begin{figure}[!ht]
    \centering
    \includegraphics[width=0.78\linewidth]{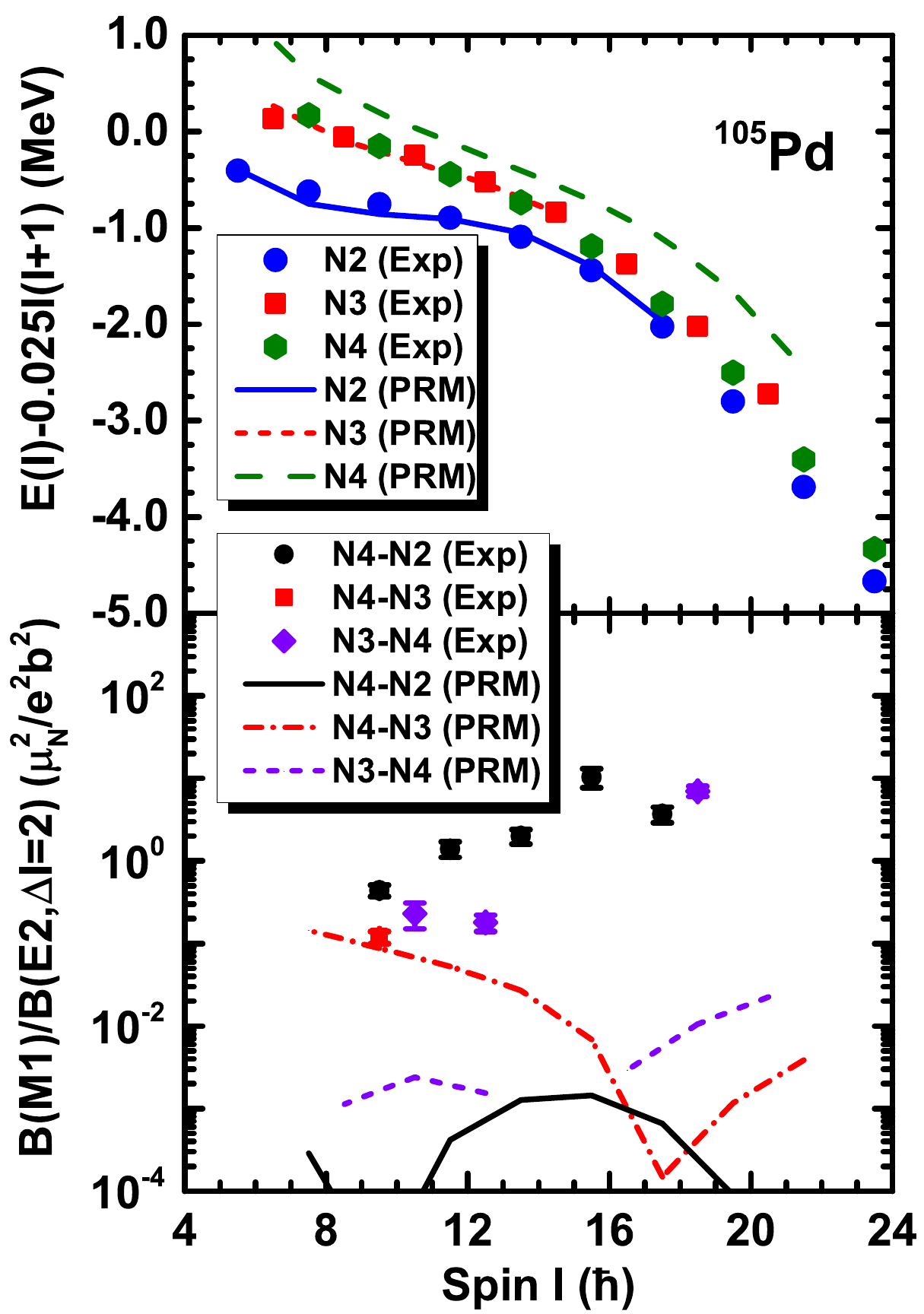}\\
    \caption{Experimental and PRM energies and $B(M1)/B(E2)$ values
    as functions of spin $I$ for band N2, N3, and N4 in $^{105}$Pd.}
    \label{fig:Energy_N4}
\end{figure}

The predicted lowest-energy negative-parity configuration 
corresponds to state C $\nu(1h_{11/2})^1$. In Ref.~\cite{105Pd} 
bands N1, N2 and N3 have been assigned to this configuration. Bands 
N1 and N2 are the unfavored and favored signature branches, while N3 
is the wobbling band corresponding to band N2. As band N4 
decays to band N2 by several transitions, one can speculate 
that this band may correspond also to this configuration 
as the two-phonon wobbling band. The fact that only 
the two lowest-energy levels of band N4 decay to band N3,
however, somewhat contradicts to this assumption. To 
check this scenario, quantum PRM calculations are performed 
for the energy spectra of bands N2, N3, and N4, as well 
as the $B(M1)/B(E2)$ ratios between these bands. The numerical 
details used in the PRM calculations are the same as in Ref.~\cite{105Pd}.
Namely, the Fermi surface locates at the beginning of the $h_{11/2}$ 
sub-shell and the empirical pairing gap is taken as 
$\Delta=12/\sqrt{A}=1.17~\textrm{MeV}$. The triaxial 
rotor is parametrized by three angular-momentum-dependent 
moments of inertia $\mathcal{J}_i=a_i\sqrt{1+bI(I+1)}$, with 
$a_{m,s,l}=5.89, 3.74, 1.27~\hbar^2/\textrm{MeV}$ and $b=0.023~\hbar^{-2}$.
The calculated values are compared with the experimental ones 
in Fig.~\ref{fig:Energy_N4}. As it is seen in the figure, 
the calculated values assuming this configuration considerably
differ from the experimental $B(M1)/B(E2)$ ratios. Thus, this 
assumption is not proved, band N4 is not the two-phonon wobbling 
band. Therefore, we cannot check from these data
that the wobbling approximation condition is met for band N3 or not.
It is worth mentioning also that none of the other
possible single-particle configurations could reproduce 
the experimental values in the calculations.

A probable scenario for the nature of band N4 is the $\gamma$ 
vibration coupled to the $\nu(1h_{11/2})^1$ orbital. Indeed, $\gamma$
bands have been previously identified in the neighboring even-even 
$^{104}$Pd~\cite{Sohler1} and $^{102}$Ru~\cite{Sohler2} nuclei.
In order to check this possibility, we have examined the decay-out
properties of band N4 to band N2. Five low-energy levels of band N4 decay 
simultaneously by $\Delta I=0$ $M1+E2$ transitions and by 
$\Delta I=2$ $E2$ transitions to the corresponding band N2 levels. 
These $\Delta I=0$ $M1+E2$ transitions have dominant $M1$ 
characters with mixing ratios around 0.4. However, if we 
derive the $B(E2;I \to I)/B(E2; I \to I-2)$ ratios to band 
N2 (the $B(E2)$ ratios of the $\Delta I=0$ and the 
$\Delta I=2$ transitions from a band N4 state to the 
subsequent two states of band N2), they are quite 
large (around 100) in spite of the small mixing ratio, 
due to the large differences in the energy and intensity 
values. This resembles the decay of the $\gamma$ vibrational 
band levels in even-even nuclei. This scenario is further confirmed
by the fact that the experimental quasiparticle alignment of 
band N4 is larger by about 2$\hbar$ than that of the band N2, as it 
is expected for the $\gamma$ band.

These observations may indicate the collective $\gamma$-band 
nature of band N4. However, it would require more investigation 
to prove this scenario. 

Recently, Jiang {\it et al.}~\cite{20} reported 
that in transitional nuclei $\beta$ vibration can cause the
doubling of the $n=0$ and $n=1$ wobbling bands. They also raise 
the possibility of such doubling in case of $\gamma$ vibration. 
Band N4 has the same signature as the yrast band, so it 
could be a good candidate for being the second $n=0$ band. 
We tried to find a band linked to band N4, which could 
be the second $n=1$ wobbling band; however, such a band could 
not be found.

Very recently, Karmakar {\it et al.}~\cite{Karmref} reported the observation
of a 254-keV and a 300-keV transitions from this band to band N1. They saw 
these $\gamma$ rays in single-gated coincidence spectra from a $\gamma$$\gamma$-coincidence matrix, which were contaminated by transitions of $^{102}$Ru and $^{104}$Pd. Based on the properties of newly observed transitions, they proposed that band N4 is a $h_{11/2}$ one-quasineutron band and band N1 is a wobbling band corresponding to it.
In our considerably cleaner double-gated coincidence spectra from a $\gamma$$\gamma$$\gamma$-coincidence cube, these transitions are not seen. Thus, we do not consider this scenario.


\subsection{Band N5}

\begin{figure}[!ht]
    \centering
    \includegraphics[width=0.78\linewidth]{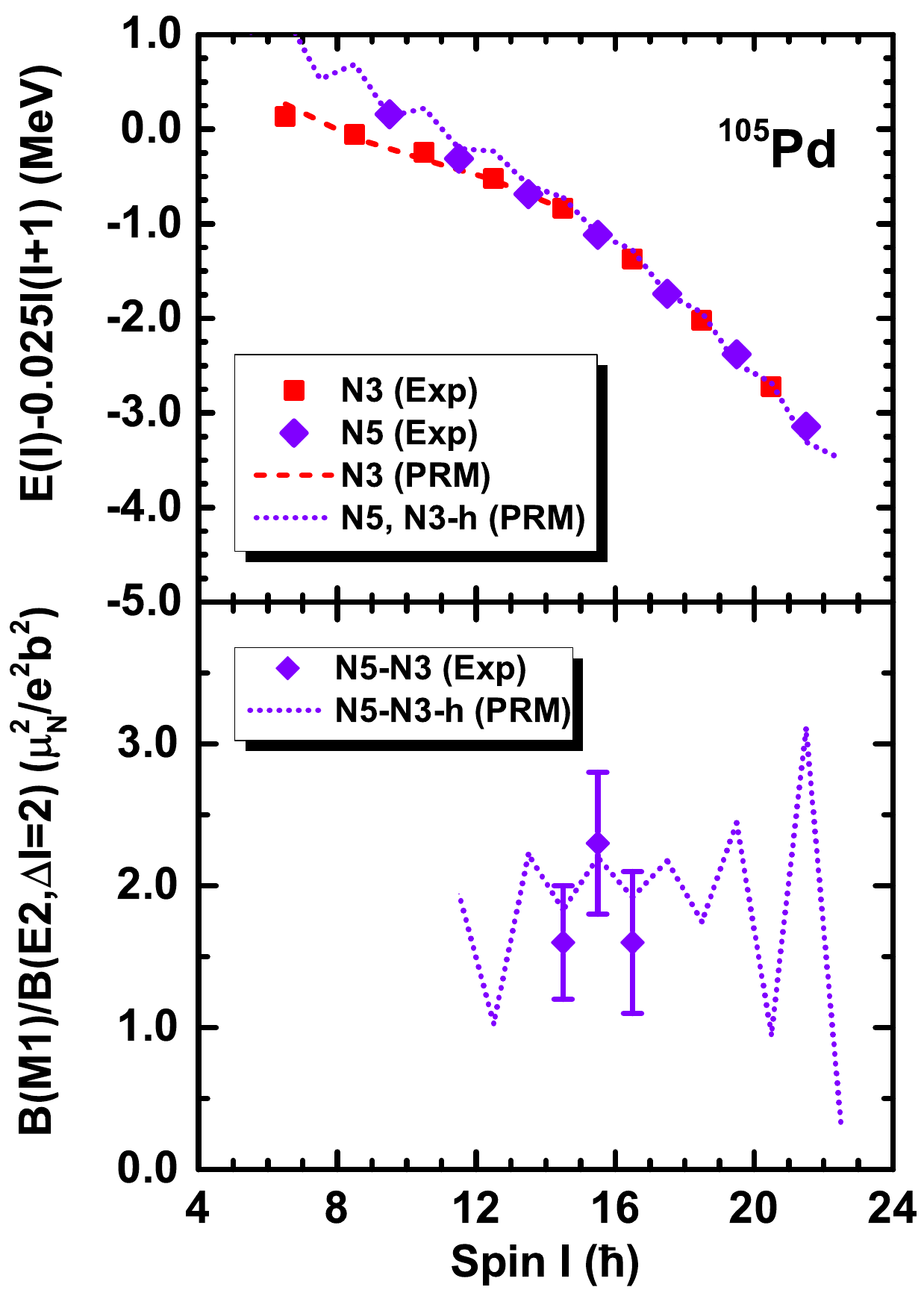}
    \caption{Experimental and PRM energies and $B(M1)/B(E2)$ 
    as a function of spin $I$ for bands N3 and N5 in $^{105}$Pd.}
    \label{fig:Energy_N5}
\end{figure}

The three-quasiparticle configuration $\pi(1g_{9/2})^{-2} \otimes \nu(1h_{11/2})^1$ 
is assigned to band N5, characterized by a signature $\alpha = -1/2$, 
and is also assigned to the high-spin part of the N3 band ($I \geq 33/2\hbar$), 
with signature $\alpha = +1/2$, as indicated in Ref.~\cite{105Pd}. In 
the PRM calculations, the deformation parameters are derived from
RDFT results. The moments of inertia are modeled using the irrotational 
flow approximation, given by $\mathcal{J}_k = \mathcal{J}_0 \sin^2(\gamma - 2k\pi/3)$
with $\mathcal{J}_0 = 21~\hbar^2/\text{MeV}$. In addition, 
one notes that the paired $\pi(1g_{9/2})^{-2}$ basis, i.e., the 
two protons occupying time-reversal states, is included
in the diagonalization of the PRM Hamiltonian. Otherwise, the obtained 
$B(M1)/B(E2)$ is too large in comparison with the experimental 
data. A comparison between the experimental and theoretical 
results reveals a good agreement in the excitation 
energies and $B(M1)/B(E2)$ transition strength ratios, as 
shown in Fig.~\ref{fig:Energy_N5}.

\begin{figure}[!ht]
\begin{center}
\includegraphics[width=\linewidth]{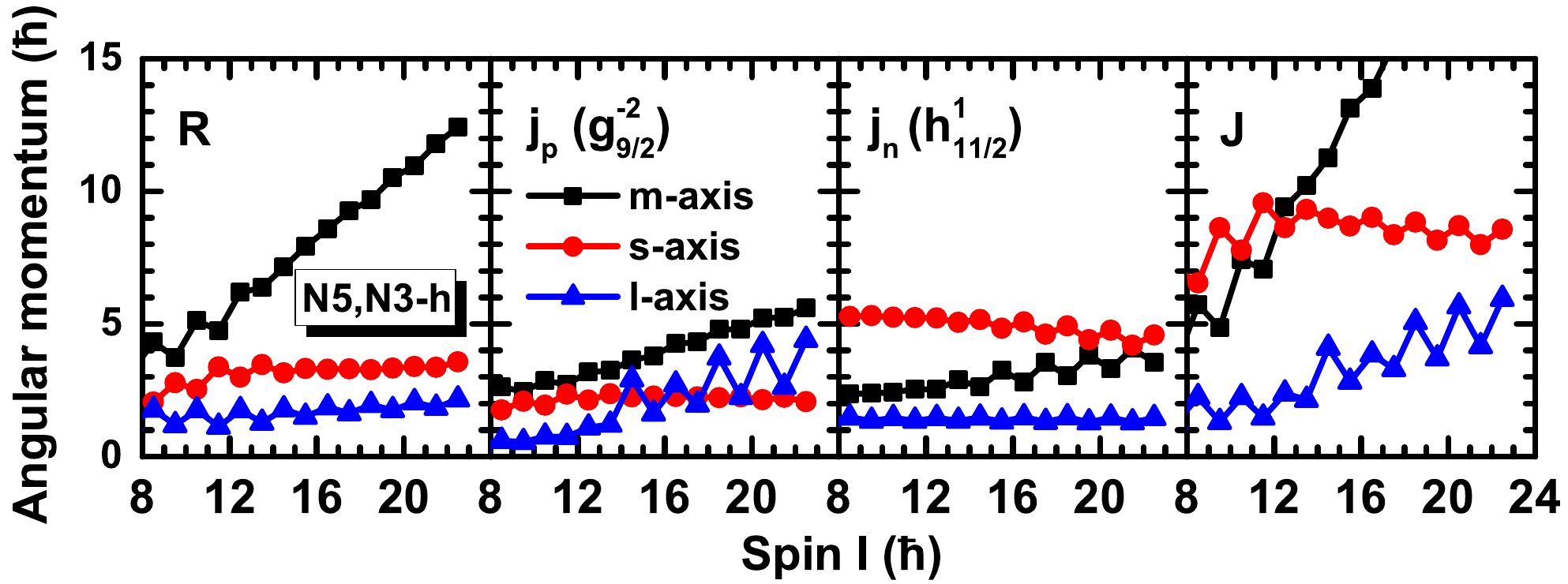}
\caption{Root mean square values of the rotor, proton,
neutron, and total angular momentum components along 
short ($s$), medium ($m$), and long ($l$) 
principal axes calculated by PRM for band N5.}
\label{fig:ANG_N5}
\end{center}
\end{figure}

In Fig.~\ref{fig:ANG_N5}, the root mean square values of 
the angular momentum components for the rotor, proton, neutron, 
and total angular momentum along the short ($s$), medium
($m$), and long ($l$) principal axes, as calculated by PRM
for band N5, are presented to examine the angular momentum geometry 
within this band. It is observed that, in the high-spin region, 
the angular momenta of the core, valence proton holes, and valence 
neutron particle are predominantly aligned along the $m$, $m$, 
and $s$ axes, respectively. As a consequence, the total angular 
momentum lies primarily in the $s$-$m$ principal plane. This 
configuration aligns with the relatively small signature 
splitting observed between band N5 and the high-spin part of
band N3. The slight odd-even staggering observed in the energy 
levels is attributed to the staggering in the $l$ axis component 
of the proton angular momentum.

\subsection{Band N6}

\begin{figure}[!ht]
    \centering
    \includegraphics[width=0.80\linewidth]{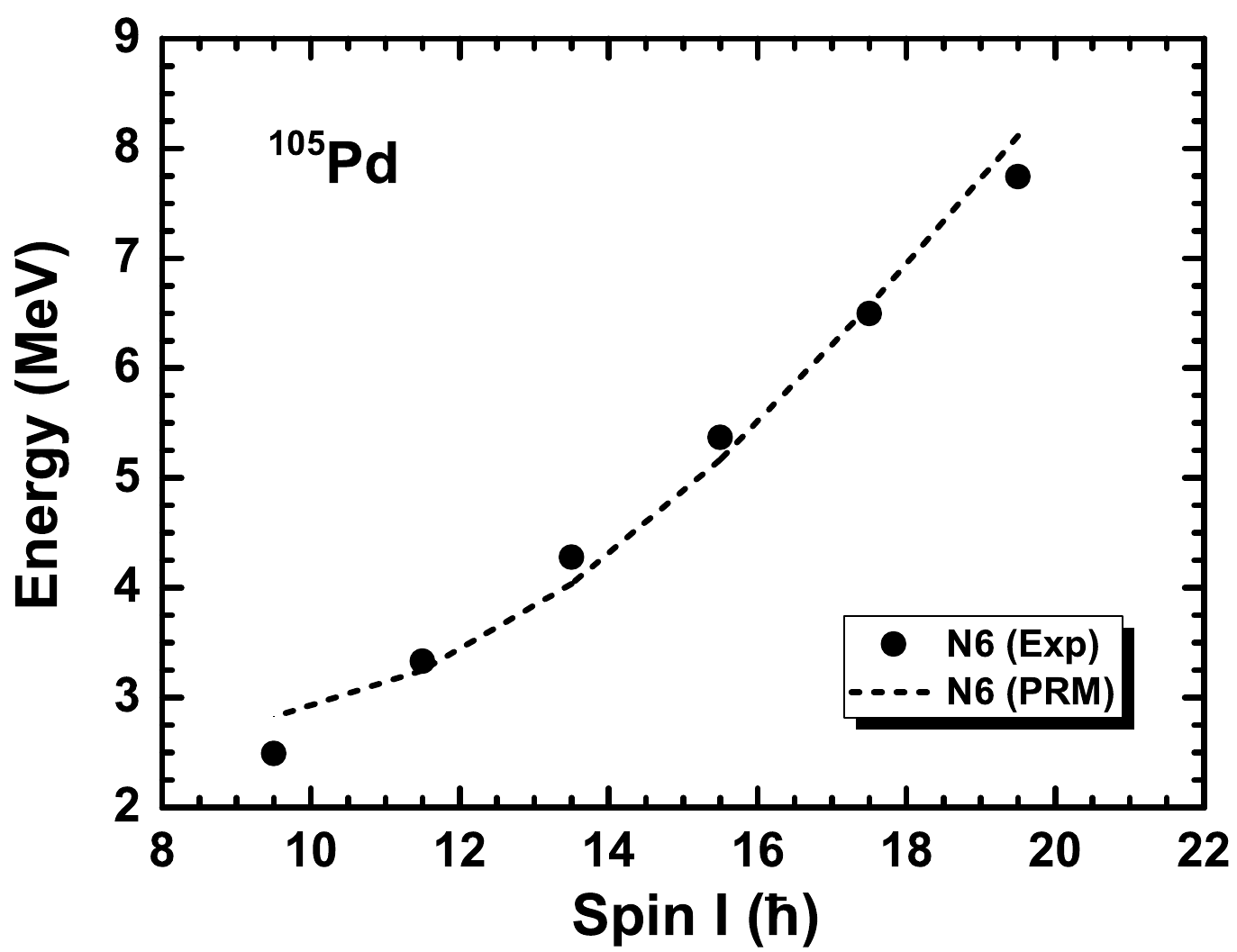}
    \caption{Experimental and PRM energy as a function of spin 
    $I$ for band N6 in $^{105}$Pd.}
    \label{fig:Energy_N6}
\end{figure}

The three-quasiparticle configuration $\nu(2d_{5/2})^2(1h_{11/2})^1$
is assigned to band N6. In the framework of PRM, a moment of 
inertia parameter $\mathcal{J}_0 = 22~\hbar^2/\text{MeV}$ is
employed for the theoretical calculations. The calculated excitation
energies derived from this model are compared with experimental 
data, as illustrated in Fig.~\ref{fig:Energy_N6}. The close 
correspondence between the theoretical predictions and the 
observed experimental values provides strong validation for 
the assigned configuration of this band. 

\begin{figure}[!ht]
\begin{center}
\includegraphics[width=\linewidth]{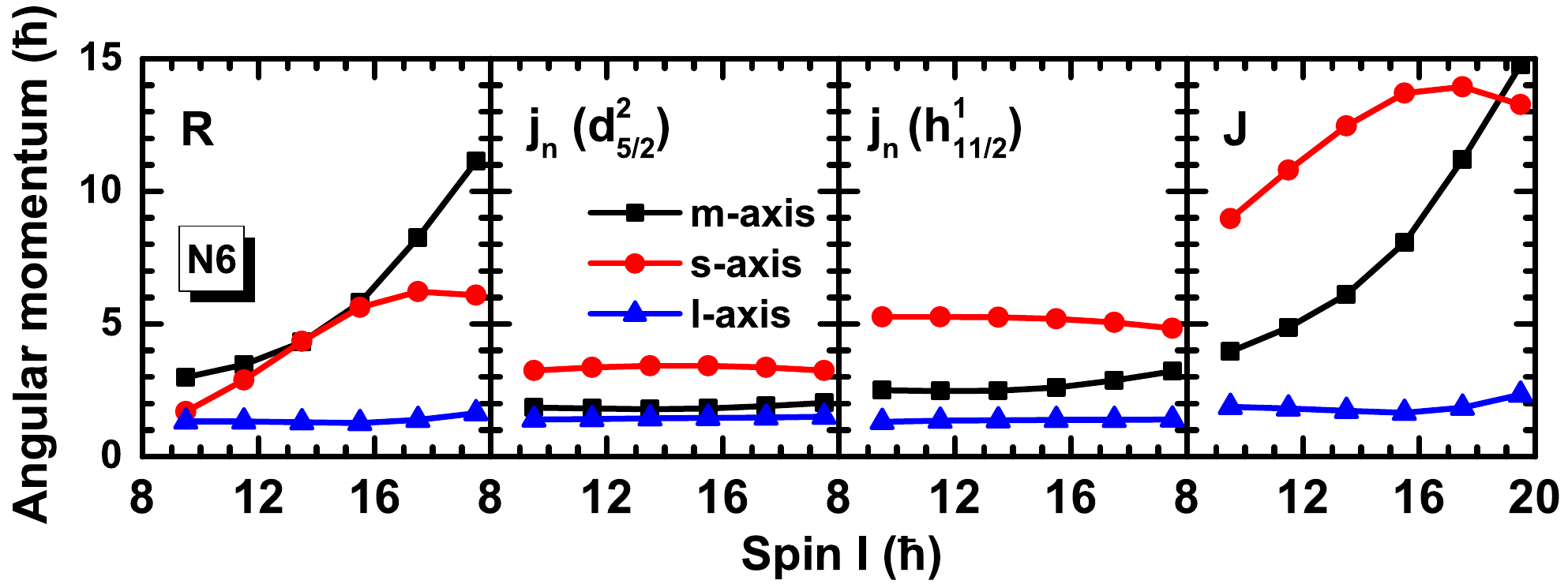}
\caption{Root mean square values of the rotor,  
neutron, and total angular momentum components along 
short ($s$), medium ($m$), and long ($l$) 
principal axes calculated by PRM for band N6.}
\label{fig:ANG_N6}
\end{center}
\end{figure}

For the rotational band N6, analysis of the root-mean-square 
values of the angular momentum components, specifically those 
of the rotor, the neutrons, and the total system, along the 
principal axes $s$, $m$, and $l$, as presented in
Fig.~\ref{fig:ANG_N6}, indicates a principal-axis
rotation predominantly aligned with the $s$ axis. This alignment 
arises due to the coherent orientation of the angular momentum 
contributions from the three neutron quasiparticles along the 
$s$ axis. Given the configuration’s notable triaxial deformation, 
quantified by a triaxiality parameter $\gamma = 20.1^\circ$, 
this band emerges as a potential candidate for a three-quasiparticle 
transverse wobbling mode. However, experimental efforts have 
thus far failed to identify its corresponding wobbling partner.

\subsection{Band N7}

\begin{figure}[!ht]
    \centering
    \includegraphics[width=0.78\linewidth]{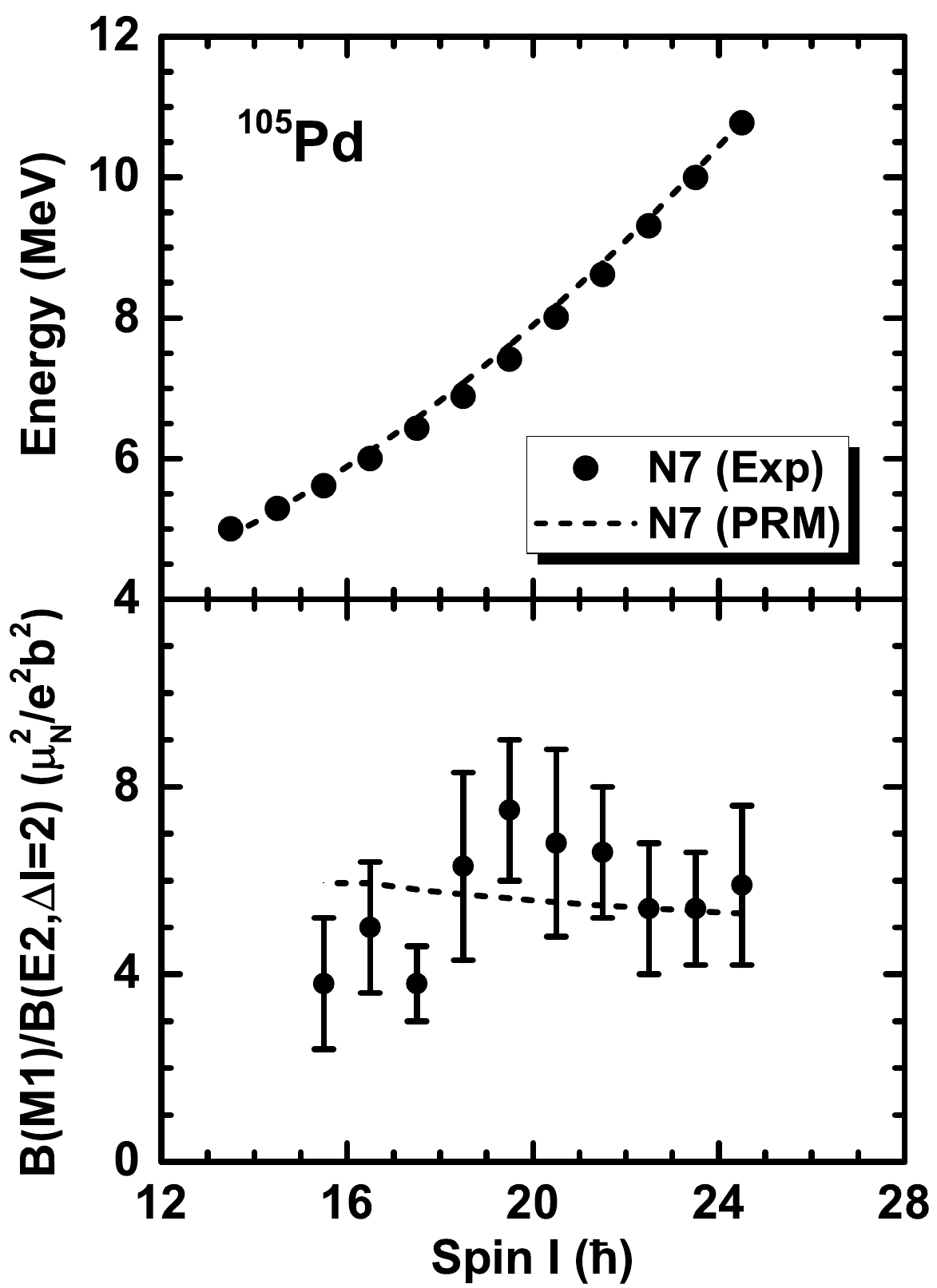}
    \caption{Experimental and PRM energy and $B(M1)/B(E2)$ 
    as functions of spin $I$ for band N7 in $^{105}$Pd.}
    \label{fig:Energy_N7}
\end{figure}

In $^{105}$Pd, experimental observations have identified 
only one strongly-coupled negative-parity band, designated 
as band N7. Theoretical calculations of the excitation 
energies and $B(M1)/B(E2)$ ratios, employing a configuration 
of $\pi(1g_{9/2})^{-2} \otimes \nu(1g_{7/2})^{-2}(1h_{11/2})^1$
and a moment of inertia parameter $\mathcal{J}_0 = 29~\hbar^2/\text{MeV}$, 
are illustrated in Fig.~\ref{fig:Energy_N7} in comparison 
with the experimental counterparts. One notes that in the 
calculations the paired $\pi(1g_{9/2})^{-2}$ basis is included
in the diagonalization of the PRM Hamiltonian. The good agreement 
between the theoretical predictions and experimental 
measurements supports the assignment of the 
$\pi(1g_{9/2})^{-2} \otimes \nu(1g_{7/2})^{-2}(1h_{11/2})^1$ 
configuration to band N7 in $^{105}$Pd.

\begin{figure}[!th]
\begin{center}
\includegraphics[width=\linewidth]{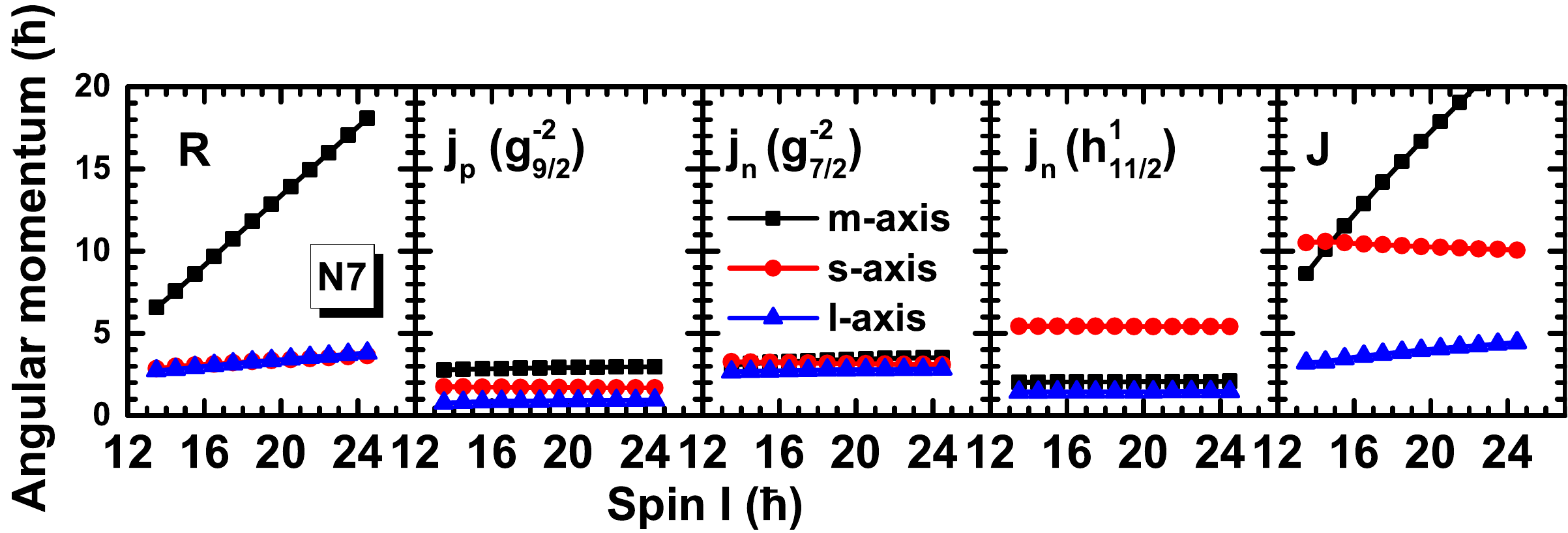}
\caption{Root mean square values of the rotor, proton, 
neutron, and total angular momentum components along 
short ($s$), medium ($m$), and long ($l$) 
principal axes calculated by PRM for band N7.}
\label{fig:ANG_N7}
\end{center}
\end{figure}

The assigned configuration is predicted to exhibit a triaxial shape,
as indicated in Table~\ref{tab:Pd105}. Notably, it includes a high-$j$ 
neutron in a particle-type state and high-$j$ protons in a hole-type 
configuration. These characteristics are considered essential for 
the manifestation of chiral rotation~\cite{1}, which has been observed 
in several neighboring nuclei. However, in the present experiment, 
no additional strongly-coupled band decaying into band N7 was observed, 
as would typically be expected in the case of chiral rotation. 
This absence raises the question of whether the chiral geometry 
is realized, specifically whether the total angular momentum lies 
outside of the principal planes. The results of 
the calculations, depicted in Fig.~\ref{fig:ANG_N7}, 
show that the angular momentum vector of the neutron $1h_{11/2}$ particle
is aligned along the $s$ axis, while the angular momentum vector 
of the two proton $1g_{9/2}$ holes is aligned along the $m$ 
axis. The reason why the two proton holes' angular momentum is 
not aligned along the $l$ axis is due to the paired $\pi(1g_{9/2})^{-2}$
basis being included. If the paired $\pi(1g_{9/2})^{-2}$ basis is 
not included in the PRM calculations, the obtained $B(M1)/B(E2)$
is about 1000 times larger than the experimental values. 
Consequently, as the rotor's angular momentum vector is also 
aligned along the $m$ axis, the total angular momentum 
vector remains confined within the principal plane defined 
by the $s$ and $m$ axes. This configuration does not satisfy 
the criteria for chiral geometry. This outcome is consistent 
with the absence of other strongly-coupled bands, further 
supporting the conclusion that chiral rotation is not realized 
in this case.

\section{Summary}

High-spin negative-parity bands of $^{105}$Pd were studied via the 
$^{96}$Zr($^{13}$C,4n) reaction using the EUROBALL IV 
$\gamma$-ray spectrometer coupled with the DIAMANT array. 
All previously reported bands were observed and extended. Moreover, 
additional new bands were identified. Altogether six negative-parity 
decoupled bands and one strongly coupled band have been assigned 
to $^{105}$Pd in the present work. 

In order to explore the configurations of the newly observed bands, 
RDFT and PRM calculations were performed. The calculated energy spectra 
and $B(M1)/B(E2)$ ratios were compared with the corresponding experimental 
values of the new bands. Based on these comparisons, quasiparticle 
configurations could be assigned to two newly observed decoupled bands 
(band N5 and band N6) and to the strongly coupled band N7. 
In band N5, the rotational mode transitions from a principal axis
rotation relative to a plane rotation with respect to the short-medium principal plane. Band N6 exhibits a principal axis rotation along the short 
axis. According to the calculated arrangement of the angular momentum vectors 
in the configuration assigned to the strongly coupled band N7, this 
configuration is not chiral, in a good agreement with the non-observation 
of other strongly coupled band in this experiment. 

The nature of band N4 could not be unambiguously revealed in this study. 
With the previous assumption that it could be the two-phonon 
wobbling band, 
the calculated and observed energies and $B(M1)/B(E2)$ ratios contradict 
to each other. 
The decay pattern of levels in band N4 to levels in band N2, as well as 
its experimental quasiparticle alignment raise the possibility that band 
N4 has a $\gamma$-vibrational character.
 
\begin{acknowledgments}

This work was supported by the National Research, Development 
and Innovation Fund of Hungary (NKFIH), financed by 
the project with contract no. TKP2021-NKTA-42 and under
the K18 funding scheme with project No. K147010, as well as by the GINOP-2.3.3-15-2016-00034 project. This 
work was also supported by the National Key R\&D Program 
of China No.~2024YFE0109803, the National Natural 
Science Foundation of China under Grants No.~12575123, 
No.~12205103, and No.~12435006, the National Key Laboratory of 
Neutron Science and Technology under Grant No.~NST202401016,
the UK STFC under grant no.~ST/P003885/1, 
and the Spanish Ministerio de Economia y Competitividad 
under Grant No. FPA2014-52823-C2-1-P and the program 
Severo Ochoa (SEV-2014-0398). B. Kruzsicz was supported 
by the PhD Excellence Scholarship from the Count István 
Tisza Foundation for the University of Debrecen, and   
by the EKÖP-24-3 University Research Scholarship Program of 
the Ministry for Culture and Innovation of Hungary from the
source of the National Research, Development 
and Innovation Fund of Hungary (NKFIH).

\end{acknowledgments}

\end{document}